# Derivation of the Generalized Time-independent Schrödinger Equation. The New Stochastic Quantum Mechanics: “Think and calculate”

**Mikhail Batanov-Gaukhman**[1]
Ph.D., Associate Professor, Institute No. 2 “Aircraft, rocket engines and power plants”,
Federal State Budgetary Educational Institution of Higher Education “Moscow Aviation Institute (National Research University)”, Volokolamsk highway 4, Moscow, Russian Federation

**Abstract:** The aim of the article is to obtain a stochastic equation that describes the averaged state of a chaotically wandering particle, regardless of its size. As a result of an analysis it was obtained the integral of the averaged action of a chaotically wandering particle in the coordinate representation. The resulting integral turned out to be a functional of the wave function $\psi(x,y,z,t)$ (92). The stochastic Euler-Poisson equations (102) was found by the calculus of variations, the solutions of which are the extremals of the functional (92). In the static case $\psi(x,y,z,t) = \psi(x,y,z)$, Eq. (102) is reduced to the generalized time-independent Schrödinger equation (113). A distinctive feature of stochastic equations (102) and (113) is the fact that they are suitable for describing the dynamics and statics of averaged states of chaotically wandering particles of any scale (i.e., they are suitable for describing stochastic objects of the microcosm and macrocosm). When solving this problem, an intermediate result was obtained: a procedure for obtaining the probability density function of an n-th order derivative for an n-fold differentiable stationary random process is defined. This result can be used in various sections of stochastic physics.

[1] alsignat@yandex.ru

## 1 BACKGROUND AND INTRODUCTION

The idea conceived by Louie de Broglie that material particles could possess wave properties was of particular importance in the 1920's. In his doctoral thesis, *Recherches sur la théorie des quanta* (Research on the Theory of the Quanta, 1924), Louis de Broglie compared the rectilinear trajectory of the free motion of a particle with a direct ray of light, and came to the conclusion that they are described by the same Jacobi equation, arising from the fundamental principle of "extremum of action". It turned out that the trajectory of the free motion of the particle and the beam of light are extrema for virtually the same functional of the action. This circumstance prompted Louis de Broglie to suggest that if the wave described by the equation

$$w = exp\{i(\omega t - \mathbf{kr})\}, \tag{1}$$

where $\omega$ is the angular frequency; $\mathbf{k}$ is the propagation vector; $t$ is time; $\mathbf{r}$ is the dimensional vector, displays some properties of a particle. The opposite assertion is quite possible that is a moving material particle can correspond to a plane wave described by

$$\psi = exp\{i(Et - \mathbf{pr})/\hbar\}, \tag{2}$$

where $E$ is the kinetic energy of a moving particle, $\mathbf{p} = m\mathbf{v}$ is its momentum, $\hbar$ is the Dirac constant (or reduced Planck constant) associated with the Planck constant by the relation $\hbar = h/2\pi$.

In addition, Louis de Broglie was acquainted with experiments, carried out by his elder brother Maurice de Broglie, which were associated with the physics of $X$-ray radiation, as well as with the pioneering work of Max Planck and Albert Einstein on the quantum nature of radiation and absorption of light. This allowed him in 1923 to 1924 to propose that a moving particle can be associated with an oscillatory perturbation $\psi$ having frequency

$$\omega = E/\hbar, \tag{3}$$

and with the wavelength

$$\lambda = 2\pi\hbar/|\mathbf{p}|. \qquad (4)$$

This idea was supported by P. Langevin and A. Einstein, but most of the physics community reacted to it with skepticism. However, in the period from 1927 to 1930, several groups of experimenters (C. Davisson & L. Germer, and O. Stern & I. Estermann et al.) showed that the idea of the existence of matter waves, proposed by de Broglie, could be used to describe the phenomenon of the diffraction of electrons and atoms in crystals.

In one of his early works of 1925 to 1926, Erwin Schrödinger, critical of the Bose-Einstein statistics formulation, wondered: "Why not start with the wave representation of the gas particles, and then impose on such ‘waves’ the quantization conditions ‘à la the Debye condition’?" After that followed his central idea: "This implies none other than the need to take seriously into consideration the proposal of L. de Broglie and A. Einstein concerning the wave theory of moving particles."

This idea served as one of the reasons that Schrödinger found the equation [1, 27 through 30]

$$i\hbar\frac{\partial\psi(\vec{r},t)}{\partial t} = -\frac{\hbar^2}{2m}\nabla^2\psi(\vec{r},t) + U(\vec{r},t)\,\psi(\vec{r},t)\,, \qquad (5)$$

where $\psi(\vec{r},t) = \psi\,(x,y,z,t)$ is the wave function describing the state of an elementary particle, $U$ is the potential energy of the particle, and $m$ is the mass of the particle.

*In fact, Erwin Schrödinger wrote the equation (4”) in "Quantisierung als Eigenwertproblem, Vierte Mitteilung", Annalen der Physik (1926) [1] in the following form*

$$\Delta\,\psi - \frac{8\pi^2}{h^2}V\,\psi \pm \frac{4\,\pi\,i}{h}\frac{\partial\psi}{\partial t} = 0. \qquad (5a)$$

*This equation acquired form (5) later.*

Schrödinger equation (5) laid the foundation for the intensive development of quantum mechanics, together with the great works of Max Planck, Albert Einstein, Niels Bohr, and Werner Heisenberg. However, the arguments presented by Schrödinger in deriving Eq. (5) were subsequently recognized by experts as incorrect, but the equation itself turned out to be correct.

This was not the only such case in science. For example, the fundamental equations of electrodynamics were derived by James Clerk Maxwell from incorrect assumptions about the mechanical properties of the ether.

Over the past ninety-five years since 1926, many researchers have proposed different ways to derive the Schrödinger equation (5) based on the axioms of many different interpretations of quantum mechanics (see for example: see for example: Rosen, N. (1964) [65]; Nelson, E. (1966) [2]; Chen, R.L.W. (1989) [38]; Vleck, V.J.H. (1994) [67]; Yung & Jick H. Yee, (1994) [68]; Peice, P. (1996) [69]; Ogiba, F. (1996) [45]; Briggs, J. & Rost, J.M., (2001) [36]; Briggs, J.S. & Rost, J.M. (2001) [70]; Hall, M. J. W. & Reginatto, M. (2002) [3]; Grössing, G. (2002) [42]; Hall, M. J.W. & Reginatto, M. (2002) [43]; Bodurov, T. (2005) [35]; Inage, S. (2006) [61]; Briggs, J., Boonchui, S. & Khemmani, S. (2007) [37]; Ward D.W., Volkmer S.M. (2008) [55]; Ricardo, C-S. (2010) [50]; Szepessy, A. (2010) [52]; Šindelka, M. (2010) [53]; Xiang-Yao W., Bai-Jun Z., Xiao-Jing L., Li-Xiao, Yi-Heng W., Yan-Wang, Qing-Cai W., Shuang C. (2011) [56]; Pranab, R. S. (2011) [49]; Field, J.H. (2011) [62]; Chiarelli, P. (2012) [39]; Schleich W.P., Greenberger D.M., Scully M.O. (2013) [64]; Sacchetti, A. (2014) [51]; Ajaib, M. A. (2015) [10]; Nanni, L. (2015) [44]; Barde, N.P., Kokne, P.M. & Bardapurkar, P. P. (2015) [34]; Wieser, R. (2015) [54]; Godart, M. (2016) [41]; Olavo, L. S. F. (2016) [47]; Baixaul, J. G. (2016) [63]; Faycal Ben Adda (2018) [57]; Chavanis, P.-H. (2018) [58]; Field, J.H. (2018) [59]; Wang, X.-S. (2018) [60]; Olavo, L. S. F. (2019) [48] and other).

A systematic review of various ways to derive the Schrödinger equation is given in the book by L. S. F. Olavo [47]. In this book, L. S. F. Olavo came to the conclusion that all the known conclusions of the Schrödinger equation can be reduced to either the *Feynman path integrals*, or to the stochastic approach, a component of which is the thermodynamic approach using the entropy of a quantum system. Olavo L. S. F. himself proposed to use the ergodic hypothesis, the central limit theorem and the stochastic Langevin equation to derive the Schrödinger equation.

But dissatisfaction in understanding the logical foundations of quantum physics remains to this day. The situation is so complicated that David Mermin suggested leaving "unnecessary disputes" and simply "Shut up and calculate!"

Nevertheless, in this article it is proposed to make one more attempt to think first and then to calculate, that is, "Think and calculate".

The probabilistic model of a randomly wandering particle (which has a volume and a continuous trajectory of motion) considered in this article clearly contradicts almost all modern interpretations of quantum mechanics, but this probabilistic model also leads to a derivation of the generalized time-independent Schrödinger equation, as indicated below.

The approach proposed in this article to the derivation of the time-independent Schrödinger equation is associated with a detailed consideration of a stationary random process in which a chaotically wandering particle of any scale (be it an electron, the nucleus of a biological cell, or the nucleus of a planet) participates. As a result of this consideration, it is possible to obtain the integral of the averaged action of the given stochastic system. As a result, the stochastic Euler-Poisson equation (102) was obtained for the extremal of the averaged action of a chaotically wandering particle. Moreover, in the case when the averaged behavior of a chaotically wandering particle does not depend on time, the stochastic Euler - Poisson equation (102) is reduced to the generalized time-independent Schrödinger equation (113). The main advantage of the obtained Eq.s

(102) and (113) is that they are suitable for describing the averaged state and behavior of stochastic systems (i.e. chaotically wandering particles) of any scale, both the microcosm and the macrocosm.

## 2 METHODS

In deriving the generalized Schrödinger equation there were applied: the methods of probability theory, the theory of stochastic processes, the theory of generalized functions, and calculus of variations. The formalism of quantum mechanics was also taken into account.

### 2.1 Probabilistic model of a particle moving along a chaotic trajectory

Consider a particle occupying a small volume compared to that of its surrounding space (see Figure 1). Conventionally, call this particle a "point".

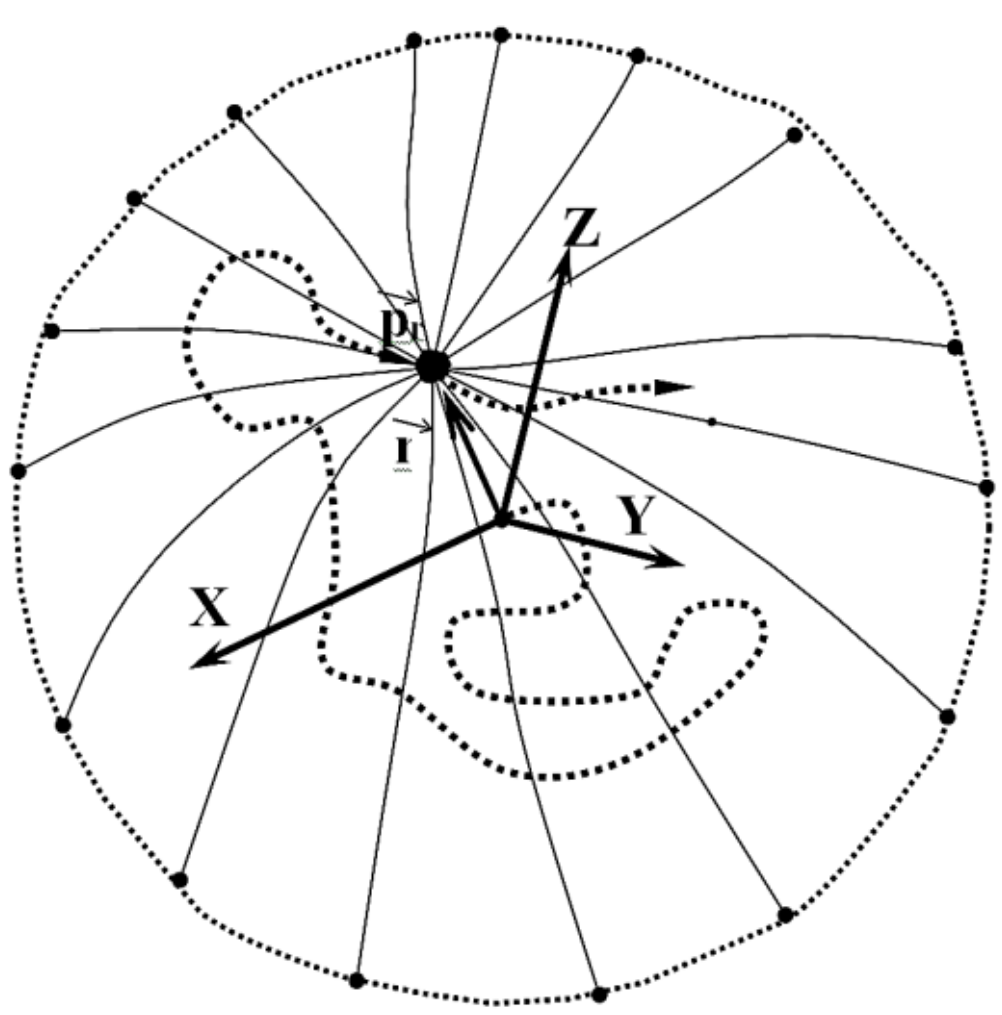

**Fig. 1:** The particle ("point") in chaotic motion in the neighborhood of the "center" of the coordinate system XYZ, so that its total mechanical energy $E$ always remains constant

Suppose that this "point" constantly chaotic motion around the conditional "center" (combined with the origin of the coordinate system XYZ) under the influence of various mutually independent force factors. Examples of such a "point" in continuous chaotic motion may be: an atom vibrating in a crystal lattice; a fly flying in a jar; a nucleus vibrating inside a biological cell; an embryo moving in the womb; a tip of a branch fluttering in the wind, and so forth.

Suppose that this “point” constantly chaotic motion around the conditional “center” (combined with the origin of the coordinate system XYZ) under the influence of various mutually independent force factors. Examples of such a "point" in continuous chaotic motion may be: an atom vibrating in a crystal lattice; a fly flying in a jar; a nucleus vibrating inside a biological cell; an embryo moving in the womb; a tip of a branch fluttering in the wind, and so forth.

We suppose that such chaotic motion of the "point" continues "forever" due to the fact that its total mechanical energy $E$ always remains constant:

$$E = T\,(x,y,z,t) \; + \; U\,(x,y,z,t) = const, \tag{6}$$

where $T(x,y,z,t)$ is the kinetic energy of the “point” due to its velocity, and $U(x,y,z,t)$ is the potential energy of the “point” associated with the force tending to return it to the “center” of the coordinate system XYZ (Figure 1).

Thus, in this model, each of the energies $T(x,y,z,t)$ and $U(x,y,z,t)$ of the "point" is a random function of time and its position relative to the "center". But these energies flow smoothly into each other so that their sum (i.e., the total mechanical energy $E$) always remains constant.

If the speed of the "point" in chaotic motion in the vicinity of the “center” (see Figure 1) is low, then according to non-relativistic mechanics, it has kinetic energy

$$T(x,y,z,t) = \frac{p_x^2(x,y,z,t) + p_y^2(x,y,z,t) + p_z^2(x,y,z,t)}{2m}\,. \tag{7}$$

For brevity, instead of (7) we write

$$T(t) = \frac{p_x^2(t) + p_y^2(t) + p_z^2(t)}{2m}, \tag{8}$$

where $p_x(t)$, $p_y(t)$, $p_z(t)$ are the respective instantaneous values of the spatial components of the momentum of the "point" in chaotic motion; $m$ is the mass of the "point".

Wherein
$$|\vec{p}(t)| = \sqrt{p_x^2(t) + p_y^2(t) + p_z^2(t)}, \tag{9}$$

where
$$p_i(t) = mv_i(t) = m\frac{dx_i(t)}{dt} = m \cdot x_i'(t). \tag{10}$$

The type of potential energy $U(x,y,z,t)$ acting on the "point" is not specified.

The action of the "point" $S$ under consideration is defined in non-relativistic mechanics as follows [12]

$$S(t) = \int_{t_1}^{t_2} [T(p_x,t) - U(x,t)]dt + Et. \tag{11}$$

To simplify the calculations, let's consider the one-dimensional case, without loss of generality. The three-dimensional case merely requires more integrations.

Due to the complexity of the path of the "point" in motion, we are interested not in the action itself (11), but rather its average over time (resp., over its realizations).

Due to the complexity of the movement of the wandering "point", we are interested not in the action itself (11), but in its average over realizations or over time

$$\overline{S} = \lim_{N\to\infty} \frac{1}{N}\sum_{i=1}^{N} S_i(t) = \int_{t_1}^{t_2} [\overline{T(p_x,t)} - \overline{U(x,t)}]dt + \overline{E}t. \tag{12}$$

Recall that for an ergodic stochastic process, an average over time is equivalent to the average over its realizations.

Finding the mean of the action (12) is carried out over the realizations, taken for the same time interval

$$\Delta t = t_2 - t_1.$$

The average kinetic energy of a "point" in chaotic motion may be represented as

$$\overline{T(p_x,t)} = \frac{1}{2m}\int_{-\infty}^{\infty} \rho(p_x)p_x^2 dp_x, \qquad (13)$$

where $\rho(p_x)$ is the probability density function of the momentum component $p_x$ of the "points".

The average potential energy of a "point" may be represented as

$$\overline{U(x,t)} = \int_{-\infty}^{\infty} \rho(x)U(x)dx, \qquad (14)$$

where $\rho(x)$ is the probability density function of the projection onto the $x$-axis of the “point” wandering in the vicinity of the conditional center (see Fig. 1 and 2).

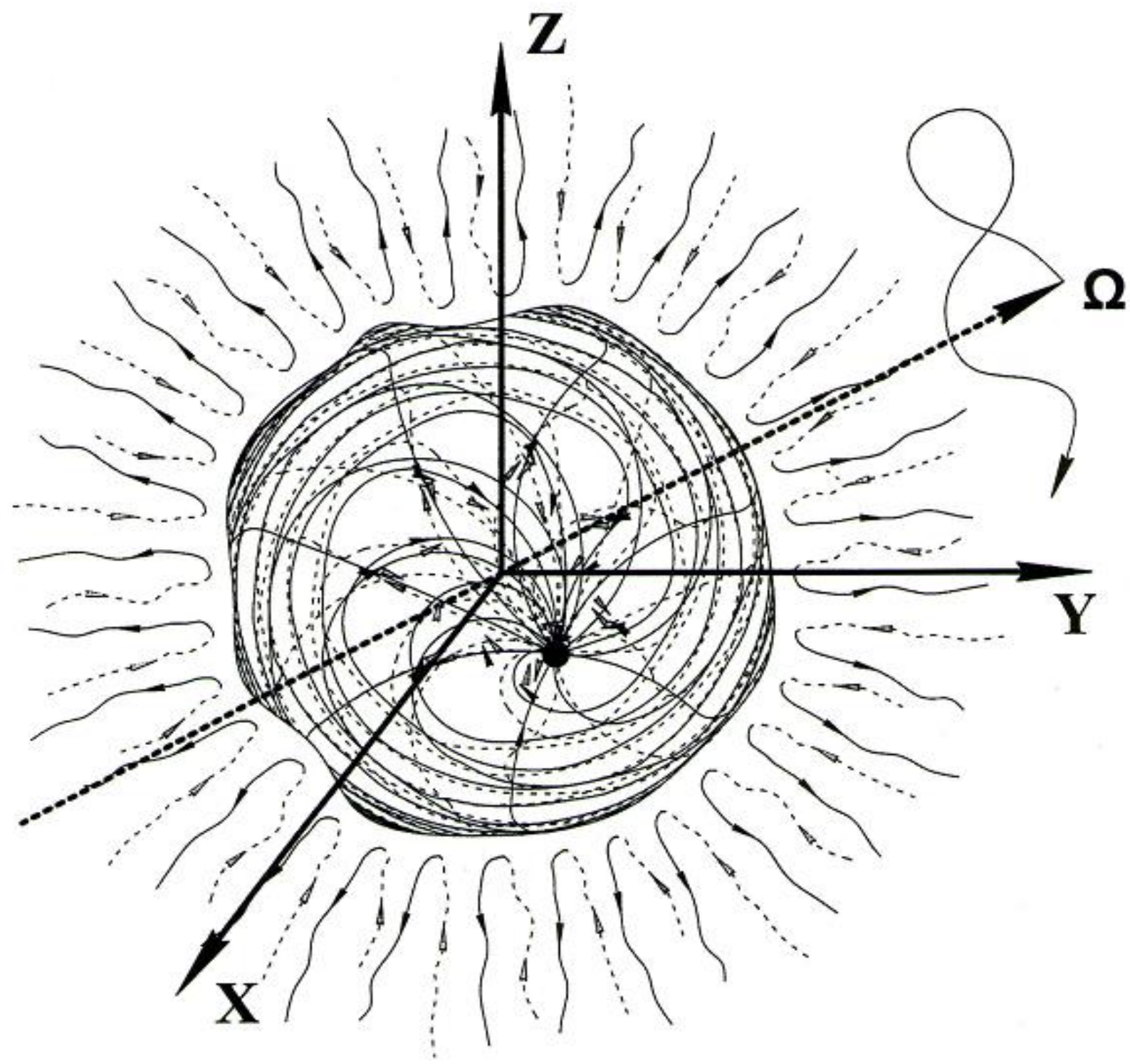

**Fig. 2:** On average, a spherically-symmetrical formation, inside which a particle (“point”) constantly randomly wanders. Ω is randomly changing axis of rotation of a spherically symmetric formation

Substituting (13) and (14) into the Eq. (12) for the mean of the action, we obtain

$$\overline{S} = \int_{t_1}^{t_2} \left\{ \frac{1}{2m} \int_{-\infty}^{\infty} \rho(p_x) p_x^2 dp_x - \int_{-\infty}^{\infty} \rho(x) U(x) dx \right\} dt + \overline{E}t. \qquad (15)$$

For the further derivation of the Schrödinger equation, two auxiliary items are given below. The first item, developed by the author of this article, is dedicated to the definition of the probability density function of the derivative of the $n$-th order of the $n$ times differentiable, stochastic stationary process. The second item, the "coordinate representation of the average momentum of a particle (i.e. “point”) is borrowed from the book of D.I. Blokhintzev [19], since this paragraph is of great importance for the aim set in this work.

## 3 RESULTS

### 3.1 Determination of the probability density function of the *n*-th derivative of an *n*-times differentiable stationary stochastic process

The key to the understanding of quantum mechanics and the limits of its application lies in the determination of the probability density function of the derivative of a stationary stochastic process, given that the probability density function of the stochastic process itself is already known.

The solution to this problem to justify the quantum-mechanical procedure of the transition from the coordinate representation to the momentum representation, and vice versa, without using the hypothesis of the existence of de Broglie waves.

This is made possible due to the fact that the momentum of a particle (material “point”) is linearly related to the derivative of its coordinate:

$$p_x = m\, \partial x/\partial t = mx'.$$

In addition, the problem of determining the one-dimensional probability density function $\rho[\xi^n(t)]$, the derivative of the $n$-th order $n$-times differentiable stationary stochastic process $\xi(t)$, when only its one-dimensional probability density function $\rho[\xi(t)]$ is known, arises in a series of problems in the fields of statistical mechanics and radio physics.

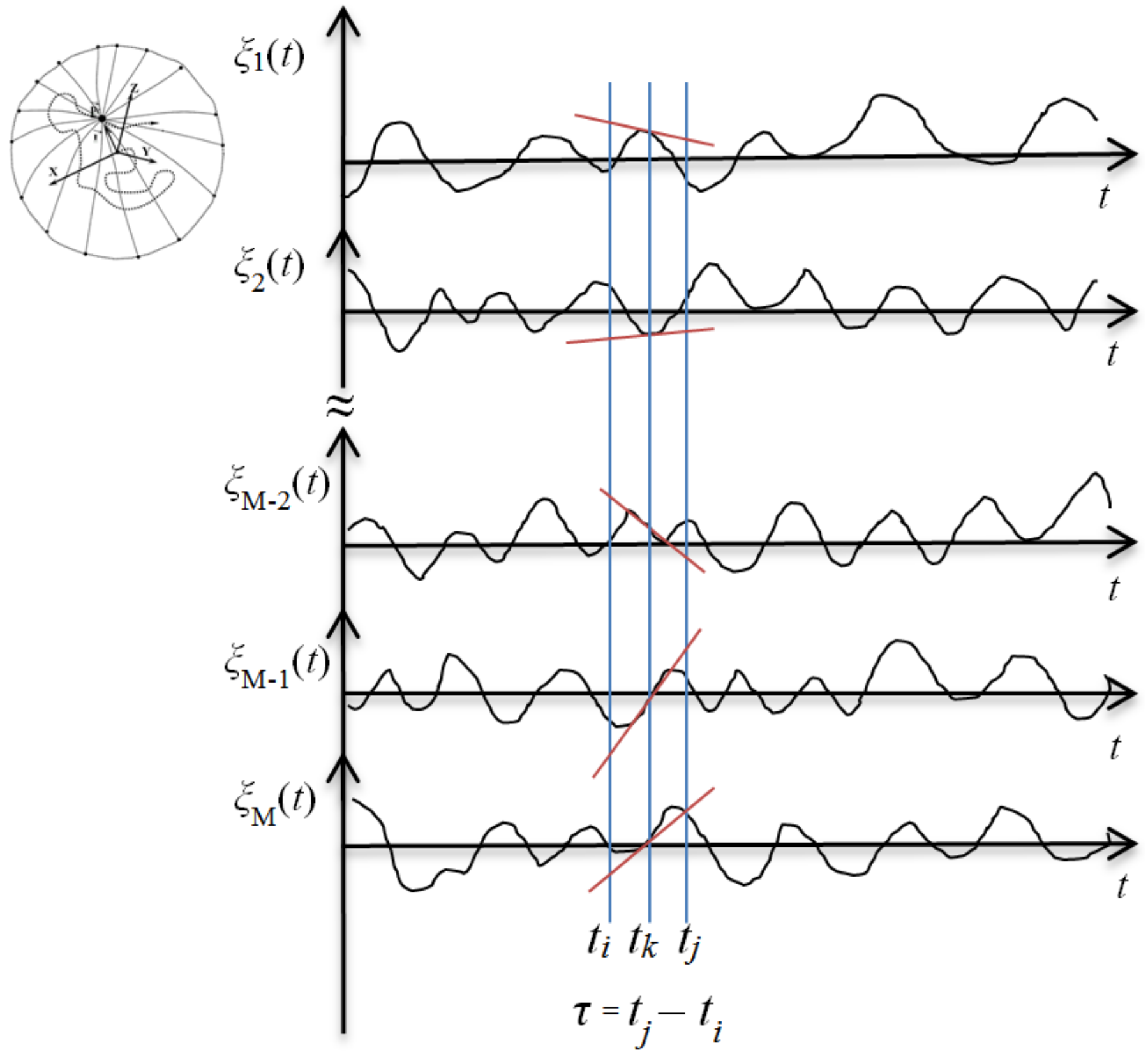

**Fig. 3:** Realizations of the differentiable stationary stochastic process $\xi(t)$. It is shown that the variables $\xi(t_k) = \xi_k$ in the cross section $t_k$ and the derivative of the stationary random process in the same cross section $\xi'(t_k) = \xi_k'$ are independent random variables

First, consider the general properties of the first derivative of the stationary stochastic process $\xi(t)$. To do this, let's explore its realizations (see Figure 3).

Figure 3 shows that the value of the random variable $\xi(t_k)$ in the cross section $t_k$ is independent from the derivative $\xi'(t_k) = \dfrac{\partial \xi(t_k)}{\partial t}$ of this process taken in the same cross section $t_k$,. Therefore, the random values $\xi(t_k)$ and $\xi'(t_k)$ are uncorrelated. This may be expressed analytically [16]:

$$< \xi(t_k)\xi'(t_k) > = < \frac{d}{dt}\frac{1}{2}[\xi(t_k)]^2 > = \frac{1}{2}\frac{d}{dt} < [\xi(t_k)]^2 > = 0, \quad (16)$$

where the brackets < > means averaging over the realizations. Here it is taken into account that the differentiation and averaging operations in this case are commutative, and that all the averaged characteristics of a stationary (in the narrow sense) stochastic process, including its dispersion, are constant over time:

$$< [\xi(t_k)]^2 > = const.$$

Realizations of a stationary stochastic process $\xi(t)$, such as that shown in Figure 3, can be interpreted as the change over time of the projection onto the *X*-axis of the position of the wandering "point" in motion (see Figures 2 and 3), i.e. $x(t) = \xi(t)$.

However, even in the case of the statistical independence of the random values $\xi(t_k) = \xi_k$ and $\xi'(t_k) = \xi_k'$, there exists a connection between the probability density functions $\rho(\xi_k)$ and $\rho(\xi_k')$. This follows from the procedure of obtaining the probability density function of a derivative $\rho(\xi'_k)$ for a known two-dimensional probability density function of a stationary stochastic process (Figures 3) [16, 17]

$$\rho(\xi_i, \xi_j) = \rho(\xi_i, t_i; \xi_j, t_j). \quad (17)$$

For this, in Ex. (17), it is necessary to make a change of variables

$$\xi_i = \xi_k - \frac{\tau}{2}\xi_k'; \quad \xi_j = \xi_k + \frac{\tau}{2}\xi_k'; \quad t_i = t_k - \frac{\tau}{2}; \quad t_j = t_k + \frac{\tau}{2}, \quad (18)$$

where $\tau = t_j - t_i; \quad t_k = \dfrac{t_j - t_i}{2}$,

with the Jacobian of the transformation $[J] = \tau$. As a result, from the probability density function (17) we obtain

$$\rho(\xi_k, \xi_k') = \lim_{\tau\to 0} \tau\, \rho\left(\xi_k - \frac{\tau}{2}\xi_k',\ t_k - \frac{\tau}{2};\ \xi_k + \frac{\tau}{2}\xi_k',\ t_k + \frac{\tau}{2}\right). \quad (19)$$

Further, integrating the obtained expression over $\xi_k$, find the desired probability density function of the derivative of the original process in the cross section $t_k$ [16, 17]:

$$\rho(\xi_k') = \int_{-\infty}^{\infty} \rho(\xi_k, \xi_k')\, d\xi_k . \tag{20}$$

The formal procedure given by (17) through (20) solves the problem of determining the probability density function $\rho(\xi')$ for a known two-dimensional probability density function (17). However, a two-dimensional probability density function is defined only for a very limited class of stochastic processes. It is therefore necessary to consider the possibility of obtaining a probability density function $\rho(\xi')$ for a known one-dimensional probability density function $\rho(\xi)$.

To solve this problem, use the following properties of stochastic processes:

1] A two-dimensional probability density function of any stochastic process can be represented as [16, 17]

$$\rho(\xi_i, t_i; \xi_j, t_j) = \rho(\xi_i, t_i)\rho(\xi_j, t_j / \xi_i, t_i), \tag{21}$$

where $\rho(\xi_j, t_j/\xi_i, t_i)$ is conditional probability density function.

2] For the strictly stationary stochastic process, the following identity holds [16, 17]

$$\rho(\xi_i, t_i) = \rho(\xi_j, t_j). \tag{22}$$

3] The conditional probability density function $\rho(\xi_j, t_j/\xi_i, t_i)$ of a stationary stochastic process as $t_i$ tends to $t_j$ degenerates into a delta function [17]

$$\lim_{\tau \to 0} \rho(\xi_j, t_j / \xi_i, t_i) = \delta(\xi_j - \xi_i). \tag{23}$$

Based on the above properties, consider a stochastic process over the interval [$t_i = t_k - \tau/2$; $t_j = t_k + \tau/2$] (see Figures 3) as $\tau$ tends to zero, using the following formal procedure. The probability density functions $\rho(\xi_i) = \rho(\xi_i, t_i)$ and $\rho(\xi_j) = \rho(\xi_j, t_j)$ can always be represented as the product of two functions:

$$\rho(\xi_i) = \varphi(\xi_i)\,\varphi(\xi_i) = \varphi^2(\xi_i),$$

$$\rho(\xi_j) = \varphi(\xi_j)\,\varphi(\xi_j) = \varphi^2(\xi_j), \tag{24}$$

where $\varphi(\xi_i)$ represents the wave function of a random variable $\xi_i$ in the cross section $t_i$. (see Figures 3); $\varphi(\xi_j)$ represents the wave function of a random variable $\xi_j$ in the cross section $t_j$.

For a strictly stationary stochastic process, we have the identity

$$\varphi(\xi_i) = \varphi(\xi_j)\,, \tag{25}$$

as is easily seen by taking the square root of both sides of the identity (22). Then, according to (24), we obtain (25). Note that identity (25) is approximately true for the majority of non-stationary stochastic processes as $\tau$ tends to zero, that is,

$$\varphi(\xi_i, t_i) = \lim_{\tau \to 0} \varphi(\xi_j,\, t_j = t_i + \tau). \tag{26}$$

When the condition (25) holds, Eq. (21) can be represented in the symmetric form

$$\rho(\xi_i, \xi_j) = \varphi(\xi_i)\rho(\xi_j / \xi_i)\,\varphi(\xi_j)\,, \tag{27}$$

where $\rho(\xi_j /\xi_i)$ is the conditional probability density function.

In expanded form (27) becomes

$$\rho\left[\xi_i, t_i = t_k - \frac{\tau}{2}; \xi_j, t_j = t_k + \frac{\tau}{2}\right] =$$

$$= \varphi\left[\xi_i, t_i = t_k - \frac{\tau}{2}\right]\rho\left[\xi_j, t_j = t_k + \frac{\tau}{2} / \xi_i, t_i = t_k - \frac{\tau}{2}\right]\varphi\left[\xi_j, t_j = t_k + \frac{\tau}{2}\right]. \tag{28}$$

In (28), let $\tau$ tend to zero, but in such a way that the interval $\tau$ is uniformly shrinks at the time $t_k = (t_j - t_i)/2$, then, taking into account (23), from (27) we obtain

$$\lim_{\tau \to 0} \rho(\xi_i, \xi_j) = \lim_{\tau \to 0} \left\{\varphi(\xi_i)\rho(\xi_j / \xi_i)\varphi(\xi_j)\right\} = \varphi(\xi_{ik})\delta(\xi_{jk} - \xi_{ik})\varphi(\xi_{jk}), \tag{29}$$

where $\xi_{ik}$ is the result of the stochastic value $\xi(t_i)$ tending to the stochastic value $\xi(t_k)$ on the left, while $\xi_{jk}$ is the result of the stochastic value $\xi(t_j)$ tending to the stochastic value $\xi(t_k)$ on the right.

Integrating both sides of the Ex. (29) over $\xi_{ik}$ and $\xi_{jk}$, we obtain

$$\int_{-\infty}^{\infty}\int_{-\infty}^{\infty}\varphi(\xi_{ik})\delta(\xi_{jk}-\xi_{ik})\varphi(\xi_{jk})d\xi_{ik}d\xi_{jk}=1. \qquad (30)$$

Ex. (30) is a formal mathematical identity out of the theory of generalized functions, taking into account the properties of the delta-function (or $\delta$-function). In order to assign the Ex. (30) a physical meaning, it is necessary to specify the specific type of the $\delta$-function.

Therefore now determine the form of a $\delta$-function for a Markov stochastic process. Consider a continuous stochastic Markov process which satisfies the Einstein - Fokker - Planck equation [17, 18]

$$\frac{\partial\rho(\xi_j/\xi_i)}{\partial t}=B\frac{\partial^2\rho(\xi_j/\xi_i)}{\partial\xi^2}, \qquad (31)$$

where $B$ is the diffusion coefficient. This parabolic differential equation has three solutions, one of which can be represented as [17, 18]

$$\rho\left(\xi_j,t_j/\xi_i,t_i\right)=\frac{1}{2\pi}\int_{-\infty}^{\infty}\exp\{iq(\xi_j-\xi_i)-q^2B(t_j-t_i)\}dq, \qquad (32)$$

where $q$ is the generalized parameter. As $t_j - t_i = \tau$ tends to zero ($\tau \to 0$), then from (32) we obtain one of the definitions of a $\delta$-function

$$\lim_{\tau\to\pm0}\rho(\xi_j/\xi_i)=\frac{1}{2\pi}\int_{-\infty}^{\infty}\exp\{iq(\xi_{jk}-\xi_{ik})\}dq=\delta(\xi_j-\xi_i). \qquad (33)$$

Since this result was obtained for the limiting case as $\tau$ tends to zero ($\tau \to 0$), it is not excluded that the $\delta$-function (33) can correspond not only to a Markov stochastic process, but also to many other stationary stochastic processes.

Substituting the $\delta$-function (33) into (30) yields

$$\int_{-\infty}^{\infty}\int_{-\infty}^{\infty}\varphi(\xi_{ik})\left[\frac{1}{2\pi}\int_{-\infty}^{\infty}\exp\{iq(\xi_{jk}-\xi_{ik})\}dq\right]\varphi(\xi_{jk})d\xi_{ik}\,d\xi_{jk}=1. \qquad (34)$$

Changing the order of integration in (34), we obtain

$$\int_{-\infty}^{\infty}\left[\frac{1}{\sqrt{2\pi}}\int_{-\infty}^{\infty}\varphi(\xi_{ik})\exp\{-iq\xi_{ik}\}d\xi_{ik}\,\frac{1}{\sqrt{2\pi}}\int_{-\infty}^{\infty}\varphi(\xi_{jk})\exp\{iq\xi_{jk})d\xi_{jk}\right]dq=1. \qquad (35)$$

According to (25), for a stationary stochastic process the condition $\varphi(\xi_{ik}) = \varphi(\xi_{jk})$ is satisfied, and also from the properties of the $\delta$-function at $\tau = 0$ follows that $\xi_{ik} = \xi_{jk} = \xi_k$. Therefore, Ex. (35) takes the form

$$\int_{-\infty}^{\infty}\varphi(q)\varphi^*(q)dq=1, \qquad (36)$$

where

$$\varphi(q)=\frac{1}{\sqrt{2\pi}}\int_{-\infty}^{\infty}\varphi(\xi_k)\exp\{-iq\xi_k\}d\xi_k, \qquad (37)$$

$$\varphi^*(q)=\frac{1}{\sqrt{2\pi}}\int_{-\infty}^{\infty}\varphi(\xi_k)\exp\{iq\xi_k\}d\xi_k. \qquad (38)$$

The integrand $\varphi(q)\varphi^*(q)$ in Ex. (36) satisfies all the requirements of the probability density function $\rho(q)$ of the random variable $q$:

$$\rho(q)=\varphi(q)\varphi^*(q)=|\varphi(q)|^2. \qquad (39)$$

Now investigate the random variable $q$. First let's reconsider the solution Ex. (32). The result of the integration on the right side of this expression does not depend on the variable $q$, therefore it may be considered as a generalized frequency. However, both the physical statement of the problem as well as the mathematical formalism in the Ex. (32) impose on $q$ the following restrictions:

1) The variable $q$ must be stochastic.

2) The random variable $q$ must characterize a stochastic process in the interval under investigation [$t_i = t_k - \tau/2$; $t_j = t_k + \tau/2$] (see Figure 3) as $\tau$ tends to zero;

3) The random variable $q$, according to the mathematical notation on the right side of (32), must belong to the set of real numbers ($q \in R$), having the cardinality of the continuum; that is, it must have the possibility to take any value in the range $]-\infty, \infty[$.

All these requirements are satisfied by any of the following random variables associated with a stochastic process in the studied time interval $\tau$:

$$\xi_i' = \frac{\partial \xi_k}{\partial t},\ \xi_i'' = \frac{\partial^2 \xi_k}{\partial t},\ \dots,\ \xi_i^{(n)} = \frac{\partial^n \xi_i}{\partial t^n}. \tag{40}$$

But these random variables do not equally characterize the process. Consider one of the realizations of the test process. The function $\xi(t)$ (see Figure 3) in the interval $\tau = t_j - t_i$ (for $\tau < \tau_{cor}$, where $\tau_{cor}$ is the correlation time of a stochastic process) may be expanded as a Maclaurin series:

$$\xi(t_j) = \xi(t_i) + \xi'(t_i)\tau + \frac{\xi''(t_i)}{2}\tau^2 + \dots + \frac{\xi^{(n)}(t_i)}{n!}\tau^n + \dots \tag{41}$$

Rewrite the Ex. (41) in the form

$$\frac{\xi_j - \xi_i}{\tau} = \xi_i' + \frac{\xi_i''}{2!}\tau + \dots + \frac{\xi_i^{(n)}\tau^{n-1}}{n!} + \dots \tag{42}$$

where $\xi(t_i) = \xi_i$, $\xi(t_j) = \xi_j$.

As in (33), let $\tau$ tend to zero, whereby (42) reduces to the identity

$$\lim_{\tau \to 0} \frac{\xi_j - \xi_i}{\tau} = \xi_k', \quad \text{where } \xi_k = \xi(t_k) \text{ (see Figure 3).} \tag{43}$$

In this way, the only random variable satisfying all the above-mentioned requirements in the interval under investigation [$t_i = t_k - \tau/2$; $t_j = t_k + \tau/2$] as $\tau$ tends to zero is the first derivative of the original stochastic process $\xi'_k$ in the cross section $t_k$. Therefore we may assume that the random variable $q$ in Ex.s (32) through (39) is directly proportional to $\xi'_k$, that is

$$q = \frac{\xi_k'}{\eta}, \tag{44}$$

where $1/\eta$ is the proportionality dimensional coefficient.

<u>An additional argument</u>: *each exponent, for example, from integral (37), corresponds to a harmonic function with frequency q*

$$\exp\{-iq\xi(t)\} \rightarrow \xi_k(t) = A\sin(qt), \tag{44a}$$

*this is one of the harmonic components of the random process ξ (t). In this case, each frequency q corresponds to the tangent of the angle of inclination of the tangent line to the harmonic function with a given frequency (see Figure 3a and Figure 3), that is,* $q \sim tg\alpha = \xi'(t)$. *Indeed, differentiating the harmonic function (44a), we obtain an unambiguous relation* $\xi_k'(t) = qA\cos(qt)$, *from which follows*

$$q = \lim_{t \to 0} \frac{\xi_k'}{A\cos(qt)} = \frac{\xi_k'}{A}. \tag{44б}$$

*For A = η, expressions (44) and (44b) coincide.*

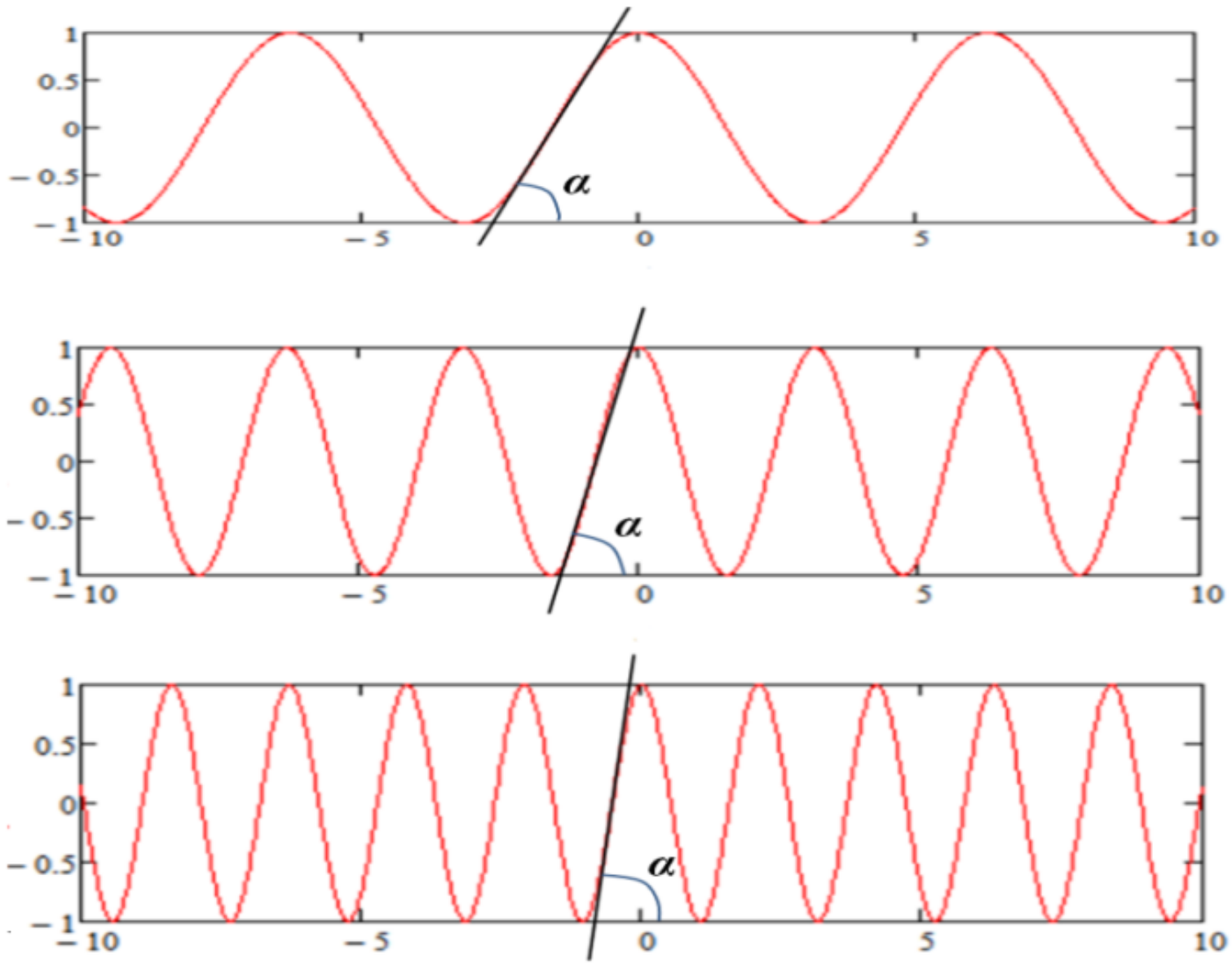

**Fig. 3a:** The higher the frequency of the harmonic function, the greater the angle α between the tangent to this function and the *t*-axis

Substituting (44) into Ex.s (36) through (39), obtain the following procedure as required for determining the probability density function of the derivative $\rho(\xi'_k)$ of a stationary stochastic (not only Markov) process $\xi(t)$ in the cross section $t_k$, for a given the one-dimensional probability density function $\rho(\xi_k)$ in the same cross section $t_k$.

1] Express the given one-dimensional probability density function $\rho(\xi)$ as the product of two wave functions $\varphi(\xi)$:

$$\rho(\xi) = \varphi(\xi)\varphi(\xi). \tag{45}$$

2] Two Fourier transforms are then carried out:

$$\varphi(\xi') = \frac{1}{\sqrt{2\pi}} \int_{-\infty}^{\infty} \varphi(\xi) \exp\{ i\xi'\xi / \eta \} d\xi\,, \tag{46}$$

$$\varphi * (\xi') = \frac{1}{\sqrt{2\pi}} \int_{-\infty}^{\infty} \varphi(\xi) \exp\{ -i\xi'\xi / \eta \} d\xi\,. \tag{47}$$

3] Finally, for any given the cross section $t$ of a stationary stochastic (not only Markov) process get the desired derivative of the probability density function:

$$\rho(\xi') = \varphi(\xi')\varphi^*(\xi') = |\varphi(\xi')|^2. \tag{48}$$

As we have already remarked, the procedure given by (45) through (48) can be applied not only to the stationary Markov processes, but to many other stationary stochastic processes for which the $\delta$-function in (30) takes the form (33).

To clarify the physical meaning of the proportionality coefficient $1/\eta$, we use a comparison with known results. This method is not mathematically perfect, but allows us to quite efficiently obtain a reliable result of practical importance.

Consider a stationary Gaussian stochastic process $\xi(t)$. Wherein, in each cross section of this process, the random variable $\xi$ is distributed according to the Gaussian distribution:

$$\rho(\xi) = \frac{1}{\sqrt{2\pi\sigma_{\xi}^{2}}} \exp\left\{-(\xi - a_{\xi})^{2}/2\sigma_{\xi}^{2}\right\}, \tag{49}$$

where $\sigma_{\xi}^{2}$ and $a_{\xi}$ are the variance and expected value of the given process $\xi(t)$.

Subjecting the probability density function (49) to the sequence of operations (45) through (48), we obtain the probability density function of the derivative of the stochastic process under consideration:

$$\rho(\xi') = \frac{1}{\sqrt{2\pi[\eta/2\sigma_{\xi}]^{2}}} \exp\left\{-\frac{\xi'^{2}}{2[\eta/2\sigma_{\xi}]^{2}}\right\}. \tag{50}$$

On the other hand, using the well-known procedure (17) through (20) for a similar case, we obtain [16, 17]

$$\rho(\xi') = \frac{1}{\sqrt{2\pi\sigma_{\xi'}^{2}}} \exp\left\{-\xi'^{2}/2\sigma_{\xi'}^{2}\right\}, \tag{51}$$

where $\sigma_{\xi'} = \sigma_{\xi}/\tau_{cor}$, and $\tau_{cor}$ is the correlation time of the stochastic process $\xi(t)$.

Comparing Ex.s (50) and (51), we find that for

$$\eta = \frac{2\sigma_{\xi}^{2}}{\tau_{cor}} \tag{52}$$

these probability density functions are completely the same.

Ex. (52) was obtained for a Gaussian stochastic process, but $\sigma_{\xi}$ is the standard deviation and correlation time $\tau_{cor}$ are the main characteristics of any stationary random process. All other initial and central moments in the case of a non-Gaussian distribution of the random variable $\xi(t)$ will give a small (insignificant) contribution to Ex. (52); therefore, it can be stated with a high degree of certainty that Ex. (52) is applicable to a wide class of stationary stochastic processes.

It should be noted that in statistical physics and quantum mechanics, the transition from the coordinate representation of a function of an elementary particle state to its momentum representation is effected by a formal process almost completely analogous to the procedure (45) through (48). The difference is only in determining the proportionality coefficient $1/\eta$.

In quantum mechanics it is well known that if the projection onto the $x$-axis of the position of a free elementary particle (for example, an electron) is described by a Gaussian distribution [13]

$$\rho(x) = |\psi(x)|^2 = \frac{1}{\sqrt{2\pi\sigma_x^2}} \exp\left\{-\frac{x^2}{2\sigma_x^2}\right\}, \tag{53}$$

where $\sigma_x$ is the standard deviation of the projections of the positions of an elementary particle onto the $x$-axis in the neighborhood of the mean (that is, the "center" of the system). Then, as the result of operations analogous to that of (45) through (48), it turns out that the probability density function of the momentum component $p_x$ of an elementary particle is also Gaussian [13]

$$\rho(p_x) = |\psi(p_x)|^2 = \frac{1}{\sqrt{2\pi\sigma_{p_x}^2}} \exp\left\{-\frac{p_x^2}{2\sigma_{p_x}^2}\right\} \tag{54}$$

with the standard deviation

$$\sigma_{p_x} = \frac{\hbar}{2\sigma_x}, \tag{55}$$

where $\hbar = 1.055 \times 10^{-34}$ J/Hz is the reduced Planck constant (or Dirac constant), which is related to the Planck constant $h = 6.626\ 070\ 15 \times 10^{-34}$ J/Hz by the ratio $\hbar = h/2\pi$.

If now take into account that the momentum component of an elementary particle (e.g. an electron) $p_x$ is equal to

$$p_x = m_e \frac{dx}{dt} = m_e x', \tag{56}$$

where $m_e$ is the electron rest mass, then taking into account (55), the probability density function (54) becomes

$$\rho(x') = \frac{1}{\sqrt{2\pi[\hbar/(m_e 2\sigma_x)]^2}} \exp\left\{-\frac{x'^2}{2[\hbar/(m_e 2\sigma_x)]^2}\right\}. \qquad (57)$$

Comparing (50) to (57), while taking into consideration (52) and that $\xi' = x'$ and $\sigma_\xi = \sigma_x$, find that for the given case

$$\eta = \frac{2\sigma_x^2}{\tau_{ex}} = \frac{\hbar}{m_e}, \qquad (58)$$

where $$\tau_{ex} = \frac{2m_e\sigma_x^2}{\hbar} = \frac{2\times 0.91\times 10^{-30}}{1.055\times 10^{-34}} \times \sigma_x^2 = 1.73\times 10^4 \sigma_x^2 \qquad (59)$$

is the correlation time of a stationary stochastic process, which is the result of the projection of the stochastic motion of the "point" (e.g., an electron) onto the $x$-axis near the stationary "center" of the system (see Figures 1 and 2).

From Ex. (58) it follows that the reduced Planck constant is not a fundamental physical constant, but a quantity expressed through the main averaged parameters of a stationary stochastic process

$$\hbar = \frac{2\sigma_{px}^2 m}{\tau_{px}}, \qquad (60)$$

where for a general case:

$\sigma_{px}$ is the standard deviation of the projection of a randomly moving particle (“point”) on the $x$-axis in the vicinity of the average value (i.e., the “center” of the system);

$\tau_{px}$ is the correlation time of a given stationary stochastic process;

$m$ is the mass of the particle ("points").

For a wide range of applications, the Ex. (60) is in itself very important, as is the related ratio (52), which in the general case may conveniently be represented as follows:

$$\eta_{px} = \frac{2\sigma_{px}^2}{\tau_{px}} = \frac{\hbar}{m} \text{ with a dimension of } (\text{m}^2/\text{s}). \qquad (61)$$

Note the following interim conclusions:

1] The quantum-mechanical transition from the coordinate representation to the momentum representation is not only applicable to the processes in the world of elementary particles, but also to any stationary Markov stochastic processes (and probably many other stochastic processes), both in the microcosm and in the macrocosm. For example, a tree branch, constantly moving chaotically around its middle position (the point of reference serving as the "center") by the rapidly changing direction of wind gusts, behaves similarly to elementary particles in the "potential well". The fluctuations of these movements of the branch would also have a discrete (quantum) average set of states, depending on the intensity of the wind gusts. With weak wind gusts, the branch generally fluctuates near the central reference point, in a way that the position of its tip can be described by a Gaussian distribution. With more intense gusts of wind, the tip of the branch rotates on average in a circle; with even greater gusts of wind, its tip basically describes the figure eight, etc. Depending on the strength of the wind, the tip of a branch can on average describe a discrete set of Lissajous figures. In other words, the quantum-mechanical formalism is not an exclusive feature of the microcosm; it is also applicable to the statistical description of many stochastic processes of the macrocosm.

2] The algorithm (45) through (48) of the transition from the coordinate representation (i.e., probability density function) $\rho(\xi_i)$ to the momentum representation (i.e., probability density function) $\rho(m\xi'_i)$ and vice versa was obtained with the concrete form of the $\delta$-function (33). It would be interesting to analyze what would be the result in the case of other types of $\delta$-function.

3] On the basis of the foregoing, we can obtain the probability density function $\rho(\xi_i'')$ of the second derivative of a stochastic process $\xi''(t)$. In this case, we should consider not the stochastic process $\xi(t)$ itself, but its first derivative $\xi'(t) = \partial\xi(t)/\partial t$. Then the distribution of the second derivative can be determined by the same procedure, only then instead of $\rho(\xi_i)$ in (45) through (48) it is necessary to substitute $\rho(\xi'_i)$.

Analogously, we also may obtain the probability density function $\rho(\xi_i^{(n)})$ of any derivative of $n$-times differentiable stationary stochastic process with the help of the following recursive procedure:

$$\rho(\xi^{(n-1)}) = \varphi\left(\xi^{(n-1)}\right)\varphi\left(\xi^{(n-1)}\right), \tag{62}$$

$$\varphi\left(\xi^{(n)}\right) = \frac{1}{\sqrt{2\pi}}\int_{-\infty}^{\infty}\varphi(\xi^{(n-1)})\exp\left\{-\frac{i\xi^{(n)}\xi^{(n-1)}}{\eta_{pn}}\right\}d\xi^{(n-1)}, \tag{63}$$

$$\varphi^*\left(\xi^{(n)}\right) = \frac{1}{\sqrt{2\pi}}\int_{-\infty}^{\infty}\varphi(\xi^{(n-1)})\exp\left\{\frac{i\xi^{(n)}\xi^{(n-1)}}{\eta_{pn}}\right\}d\xi^{(n-1)}, \tag{64}$$

$$\rho(\xi^{(n)}) = \varphi\left(\xi^{(n)}\right)\varphi^*\left(\xi^{(n)}\right),$$

where

$$\eta_{pn} = \frac{2\sigma^2_{\xi^{(n-1)}}}{\tau_{cor\,\xi^{(n-1)}}}, \tag{65}$$

where $\sigma^2_{\xi^{(n-1)}}$, $\tau_{cor\,\xi^{(n-1)}}$ are the variance and the correlation time, respectively, of the given $n-1$ times differentiable stationary stochastic process.

4] The procedure (45) through (48), is completely analogous to the quantum-mechanical transition from the coordinate representation of a quantum system to its momentum representation, obtained here on the basis of a study of realizations of an ordinary stochastic stationary process, i.e. without involving the phenomenological principles of wave-particle duality.

There is also no need to use the de Broglie hypothesis about the existence of matter waves to describe the diffraction of atoms and electrons on the crystal lattice. We refer to [72, arXiv:2007.13527], where, based on the laws of geometric optics and the principles of statistical physics, a formula was obtained for calculating the volumetric scattering diagrams of particles on a multilayer crystal surface:

$$\rho(\nu,\omega/\vartheta,\gamma)=\frac{1}{2\pi^2 l_2}\left(\frac{(\cos^2\pi n_1-\cos(\pi n_1)\cos\left(\sqrt{\dfrac{a^2+b^2}{d^2}}l_2/\eta\right)}{\left(\pi n_1/l_2\right)^2-\left(\sqrt{\dfrac{a^2+b^2}{d^2}}l_2/\eta\right)^2}-\frac{\cos\left(\pi n_1+\sqrt{\dfrac{a^2+b^2}{d^2}}l_2/\eta\right)-1}{\left(\pi n_1/l_2+\sqrt{\dfrac{a^2+b^2}{d^2}}/\eta\right)^2}\right)\times$$
$$\times\left|\frac{d\left(a'_\nu b'_\omega-a'_\omega b'_\nu\right)+c'_\nu\left(b a'_\omega-a b'_\omega\right)}{d^2\sqrt{a^2+b^2}}\right|, \tag{66}$$

where $\vartheta$, $\gamma$, $\omega$, $\nu$ *is* angles, are shown in Figure 4b;

$a=\cos\nu\cos\omega+\cos\vartheta\cos\gamma$; $b=\cos\nu\sin\omega+\cos\vartheta\sin\gamma$; $d=\sin\nu+\sin\vartheta$;
$a_\nu{}'=-\sin\nu\cos\omega$; $b_\nu{}'=-\sin\nu\sin\omega$; $c_\nu{}'=\cos\nu$; $a_\omega{}'=-\cos\nu\sin\omega$;
$b_\omega{}'=\cos\nu\cos\omega$,

$$\eta=\frac{l_1^2(\pi^2 n_1^2-6)}{6\pi^2 r_{cor}},$$

here $l_2 = l_1 n_1$ is the depth of the multilayer surface of the crystal is effectively involved in the scattering (reflection) of microparticles; $l_1$ is the thickness of one layer, i.e. one sinusoidal equipotential surface; $n_1$ is the number of layers effectively involved in the scattering of the microparticles; $r_{cor}$ is the average radius of curvature of a sinusoidal equipotential surface.

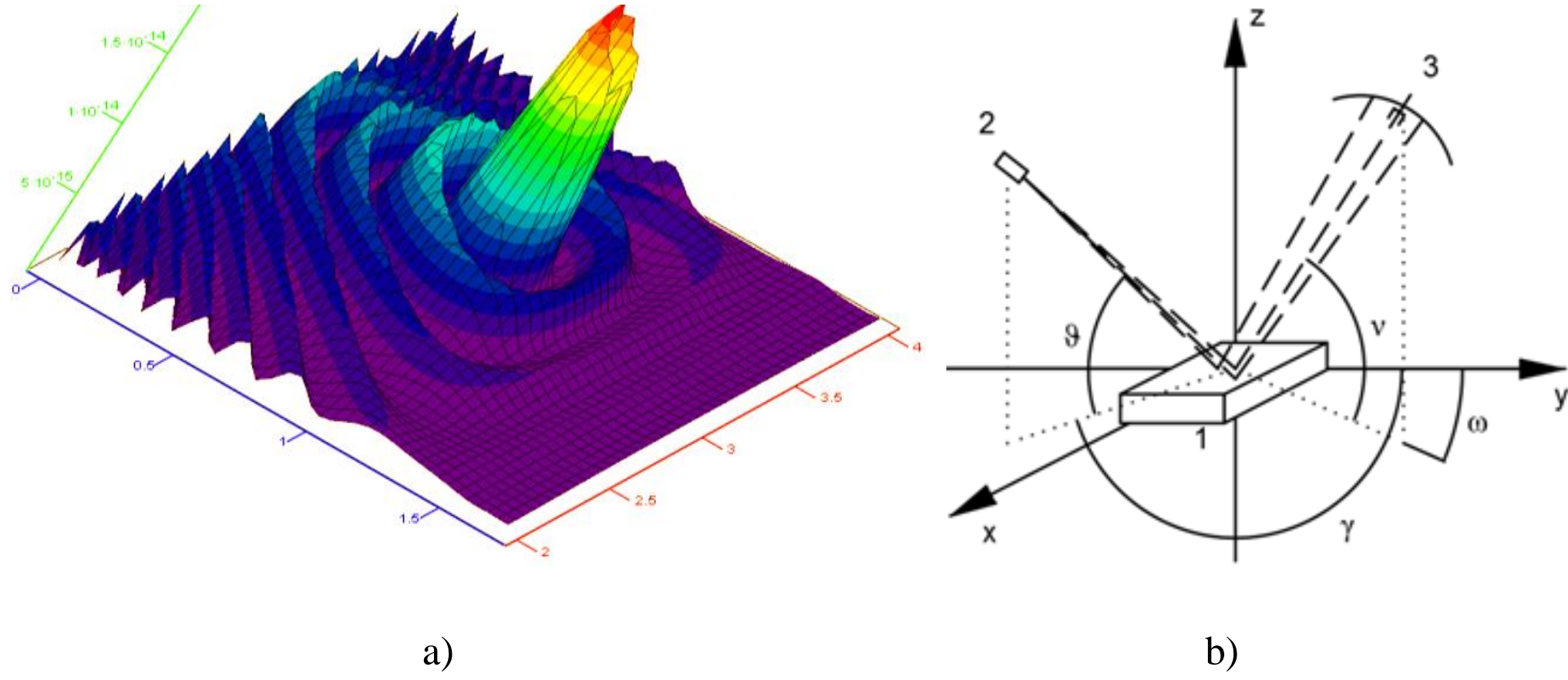

**Fig. 4:** a) Diagram of scattering of particles (electrons) on 64 layers of sinusoidal equipotential crystal surfaces, calculated according to formula (66) using MathCad software [72]; b) Diffraction of microparticles on a crystal: 1 – crystal, 2 – microparticle generator, 3 – microparticle detector

For a single crystal, all the sinusoidal equipotential surfaces have the same, that is, $r_{cor}$. Therefore in this case $r_{cor}$ signifies the effective cross section of the scattering of electrons by the atoms of a crystal.

The results of a calculation using (66) at an angle of incidence of particles on the surface of the crystal using the values $\vartheta = 45^\circ$, azimuth angle $\gamma = 0^\circ$ (see Figure 4b), $l_1 = 10^{-11}$ cm, $r_{cor} = 6\cdot 10^{-9}$ cm, $n_1 = 64$ (layers) are shown in Figure 4a.

**3.2 The coordinate representation of the average particle momentum**

The contents of this paragraph are well known to specialists in the field of quantum mechanics. However, given the importance of subsequent conclusions, the following calculations are almost completely rewritten from [19].

Let's first recall the properties of the Dirichlet integral appearing in the theory of Fourier integrals and the theory of generalized functions [19]

$$\lim_{k\to\infty}\frac{1}{\pi}\int_a^b \varphi(z)\frac{\sin kz}{z}dz == \int_a^b \varphi(z)\delta(z)dz = \begin{cases} 0, if \;\; a,b>0 \;\; or \;\; a,b<0, \\ \varphi(0), \; if \;\; a<0, \;\; b>0, \end{cases} \quad (67)$$

whereby

$$\lim_{k\to\infty}\frac{1}{\pi}\frac{\sin kz}{z} = \delta(z) \quad (68)$$

this is one of many forms of a $\delta$-function.

Now consider the case of one dimension to reduce calculations and prove the equality [19]

$$\overline{p_x^n} = \int_{-\infty}^{+\infty}\rho(p_x)p_x^n dp_x = \int_{-\infty}^{+\infty}\psi(p_x)p_x^n\psi(p_x)dp_x = \int_{-\infty}^{+\infty}\psi(x)\left(-i\hbar\frac{\partial}{\partial x}\right)^n\psi(x)dx, \quad (69)$$

where $n$ is a positive integer; $\overline{p_x^n}$ is averaging over time (or over implementations) of the momentum component raised to the power $n$

$$p_x^n = (m\cdot\partial x/\partial t)^n = (mx')^n, \quad (70)$$

where $\psi(x)$ *and* $\psi(p_x)$ are the wave functions (probability amplitude densities) which were introduced in (24) [ $\psi(x)=\varphi(x)$ ] and (48) [ $\psi(p_x)=\varphi(p_x)=\varphi(mx')$ ], and, according to (46) and (47), are related (provided the stationary stochastic process) by the Fourier transforms:

$$\psi\left(p_x = mx'\right) = \int_{-\infty}^{+\infty} \psi(x) \frac{e^{i\frac{x'x}{\eta_{par}}}}{(2\pi)^{1/2}} dx = \int_{-\infty}^{+\infty} \psi(x) \frac{e^{i\frac{p_x x}{\hbar}}}{(2\pi\hbar)^{1/2}} dx, \qquad (71)$$

$$\psi^*\left(p_x = mx'\right) = \int_{-\infty}^{+\infty} \psi(x) \frac{e^{-i\frac{x'x}{\eta_{par}}}}{(2\pi)^{1/2}} dx = \int_{-\infty}^{+\infty} \psi(x) \frac{e^{-i\frac{p_x x}{\hbar}}}{(2\pi\hbar)^{1/2}} dx, \qquad (72)$$

where the parameter $\eta_{par}$ is defined by (61)

$$\eta_{par} = \frac{2\sigma_{parx}^2}{\tau_{parx}} = \frac{\hbar}{m}. \qquad (73)$$

In order to prove statement (69), for $\psi(p_x)$ and $\psi^*(p_x)$ substitute their respective Ex.s in terms of integrals (71) and (72) [19]

$$\overline{p_x^n} = \int_{-\infty}^{+\infty} \left[ \int_{-\infty}^{+\infty} \psi(x_i) \frac{e^{-i\frac{p_x x_i}{\hbar}}}{(2\pi\hbar)^{1/2}} dx_i \, p_x^n \int_{-\infty}^{+\infty} \psi(x_j) \frac{e^{i\frac{p_x x_j}{\hbar}}}{(2\pi\hbar)^{1/2}} dx_j \right] dp_x. \qquad (74)$$

A direct check makes it easy to verify that

$$p_x^n e^{i\frac{p_x x_j}{\hbar}} = \left( -i\hbar \frac{\partial}{\partial x_j} \right)^n e^{i\frac{p_x x_j}{\hbar}}. \qquad (75)$$

Substituting (75) into (74), we obtain

$$\overline{p_x^n} = \frac{1}{2\pi} \int_{-\infty}^{+\infty} \left[ \int_{-\infty}^{+\infty} \psi^*(x_i) e^{i\frac{p_x x_i}{\hbar}} dx_i \int_{-\infty}^{+\infty} \psi(x_j) \left( i\hbar \frac{\partial}{\partial x_j} \right)^n e^{-i\frac{p_x x_j}{\hbar}} dx_j \right] dp_x. \qquad (76)$$

Let's perform an integration by parts $n$ times, starting from the second integral in the integrand. In doing so, assume that $\psi(x)$ and its derivatives vanish at the integration boundaries, that is, at $x = \pm\infty$. Following these actions, we find

$$\overline{p_x^n} = \frac{1}{2\pi\hbar} \int_{-\infty}^{+\infty} \left[ \int_{-\infty}^{+\infty} \psi(x_i) e^{-i\frac{p_x x_i}{\hbar}} dx_i \int_{-\infty}^{+\infty} e^{i\frac{p_x x_j}{\hbar}} \left( -i\hbar \frac{\partial}{\partial x_j} \right)^n \psi(x_j) dx_j \right] dp_x. \qquad (77)$$

Changing the order of integration, first integrate over $p_x$ [19]

$$\overline{p_x^n} = \frac{1}{2\pi\hbar}\int_{-\infty}^{+\infty}\psi(x_i)dx_i\int_{-\infty}^{+\infty}\left(-i\hbar\frac{\partial}{\partial x_j}\right)^n\psi(x_j)dx_j\int_{-\infty}^{+\infty}e^{i\frac{p_x(x_j-x_i)}{\hbar}}dp_x. \quad (78)$$

Introduce the variables $\zeta = p_x/h$, $z = x_j - x_i$. In the last integral in (78), we perform the integration over $\zeta$ between the limits $-k$ to $+k$; then passing to the limit $k \to \infty$, this expression takes the form

$$\begin{aligned}\overline{p_x^n} &= \int_{-\infty}^{+\infty}\left[\left(-i\hbar\frac{\partial}{\partial x}\right)^n\psi(x)\right]dx\cdot\lim_{k\to\infty}\int_{-\infty}^{+\infty}\psi(x+z)\frac{\sin kz}{\pi z}dz = \\ &= \int_{-\infty}^{+\infty}\left[\left(-i\hbar\frac{\partial}{\partial x}\right)^n\psi(x)\right]dx\int_{-\infty}^{+\infty}\psi(x+z)\delta(z)dz.\end{aligned} \quad (79)$$

Based on the properties of the Dirichlet integral (67), when $a = -\infty$; $b = +\infty$ and $\psi(z) = \psi(x+z)$, we have [19]

$$\overline{p_x^n} = \int_{-\infty}^{+\infty}\left[\left(-i\hbar\frac{\partial}{\partial x}\right)^n\psi(x)\right]\psi(x)dx = \int_{-\infty}^{+\infty}\psi(x)\left(-i\hbar\frac{\partial}{\partial x}\right)^n\psi(x)dx, \quad (80)$$

thus, Ex. (69) is proved [19].

Similarly, it can be shown that for slow changes in the wave function, i.e. for $\psi(x,t) \approx \psi(x,t+\Delta t)$, the following expressions are true for each instant of time $t$

$$\overline{p_x^n}(t) = \int_{-\infty}^{+\infty}\psi(x,t)\left(-i\hbar\frac{\partial}{\partial x}\right)^n\psi(x,t)dx, \quad (81)$$

and

$$\overline{p_x^n}(t) = \int_{-\infty}^{+\infty}\psi(x,t)\left(-i\hbar\frac{\partial}{\partial x}\right)^n\psi^*(x,t)dx, \quad (81a)$$

where $\psi(x,t) = \varphi(x,t)\exp\{iut\}$, $\psi^*(x,t) = \varphi(x,t)\exp\{-iut\}$,

here $u$ is an arbitrary real number.

It can also be shown that the $n$-th power of the averaged insignificant change with time in the total mechanical energy $\overline{dE^n}$ of a chaotically wandering particle can be expressed through the wave function $\psi(x,t) \approx \psi(x,t+\Delta t)$

$$\overline{dE^n} = \int_{-\infty}^{+\infty} \psi(x,t)\left(-i\hbar\frac{\partial}{\partial t}\right)^n \psi(x,t)dx. \qquad (82)$$

Let’s divide both sides of expression (81) by the particle mass $m$ and take into account (73), as a result, we obtain a similar statement of stochastic quantum mechanics for the average velocity of a chaotically wandering particle

$$\overline{v_x^n}(t) = \int_{-\infty}^{+\infty} \psi(x,t)\left(-i\eta_{px}\frac{\partial}{\partial x}\right)^n \psi(x,t)dx. \qquad (83)$$

The generalization to three dimensions then increases the number of integrations in Ex.s (80) through (83).

### 3.3 Derivation of the stochastic Euler-Poisson’s equation

Let's return to the average action of the particle ("point") in chaotic motion (15)

$$\overline{S} = \int_{t_1}^{t_2}\left\{\frac{1}{2m}\int_{-\infty}^{\infty}\rho(p_x)p_x^2dp_x - \int_{-\infty}^{\infty}\rho(x)U(x)dx\right\}dt + \overline{E}t.$$

First, let's answer the question: - How can the average potential energy of a chaotically wandering particle change in time, if the considered stochastic model of its behavior is based on Ex. (6)? To answer this question, we average the Ex. (6)

$$E = \overline{T(x,t)} + \overline{U(x,t)} = const. \qquad (84)$$

In this case, it becomes obvious that the averaged potential energy of the considered stochastic system can change if the kinetic energy of this system also changes in inverse proportion to it. For example, if $\overline{U(x,t)}$ increases, then $\overline{T(x,t)}$ should decrease in such a way that Ex. (84) remains in force.

Therefore, the averaged action (15) can be represented as

$$\overline{S} = \int_{t_1}^{t_2} \left\{ \frac{1}{2m} \int_{-\infty}^{\infty} \rho(p_x, t) p_x^2 dp_x - \int_{-\infty}^{\infty} \rho(x,t) U(x,t) dx \right\} dt + \overline{E} t. \qquad (85)$$

Now let's represent action (85) in coordinate form. To do this, perform the following operations:

1] Writing the probability density function $\rho(x)$ as a product of two wave functions:

$$\rho(x,t) = \psi(x,t)\,\psi(x,t). \qquad (86)$$

2] Let's use the coordinate representation of the average impulse raised to the $n$-th power (80). Wherein, in particular, for $n = 2$, we have

$$\overline{p_x^2(t)} = \int_{-\infty}^{+\infty} \rho(p_x, t) p_z^2 dp_x = \int_{-\infty}^{+\infty} \psi(x,t) \left( -i\hbar \frac{\partial}{\partial x} \right)^2 \psi(x,t) dx. \qquad (87)$$

3] Using (87), we represent the average kinetic energy of the particle (“point”) (13) in the form

$$\overline{T(x,t)} = \frac{1}{2m} \overline{p_x^2(x,t)} = \frac{1}{2m} \int_{-\infty}^{\infty} \rho(p_x, t) p_x^2 dp_x = \frac{\hbar^2}{2m} \int_{-\infty}^{\infty} \psi(x,t) \frac{\partial^2 \psi(x,t)}{\partial x^2} dx. \qquad (88)$$

4] By taking (85) into consideration, now represent the average potential energy of the particle ("point") (14) in the form:

$$\overline{U(x,t)} = \int_{-\infty}^{\infty} \psi(x,t) U(x,t) \psi(x,t) dx. \qquad (89)$$

5] In the case when the total mechanical energy of a chaotically wandering particle changes insignificantly with time, we can write

$$E(t) = E(t_0) \pm dE(t_0 + \Delta t).$$

Let's average this expression

$$\overline{E(t)} = \overline{E(t_0)} \pm \overline{dE(t_0 + \Delta t)},$$

where $\overline{dE(t_0 + \Delta t)}$ is the average insignificant change in the total mechanical energy of a chaotically wandering particle over a short time interval $\Delta t$.

Let's represent this equation, taking into account (82), in the coordinate form

$$\overline{E(t)} = \int_{-\infty}^{\infty} \psi(x,t) E \psi(x,t) dx \pm \int_{-\infty}^{\infty} i\hbar \psi(x,t) \frac{\partial \psi(x,t)}{\partial t} dx = dx. \qquad (90)$$

6] Substituting Ex.s (88), (89) and (90) into integral (85), we obtain a record of the average action of a chaotically wandering particle in the coordinate representation for the case when its total mechanical energy changes insignificantly with time

$$\overline{S} = \int_{t_1}^{t_2} \{ \frac{\hbar^2}{2m} \int_{-\infty}^{\infty} \psi(x,t) \frac{\partial^2 \psi(x,t)}{\partial x^2} dx - \int_{-\infty}^{\infty} \psi(x,t) U(x,t) \psi(x,t) dx + \int_{-\infty}^{\infty} \psi(x,t) E \psi(x,t) dx dt \pm$$

$$\pm \int_{-\infty}^{\infty} i\hbar \psi(x,t) \frac{\partial \psi(x,t)}{\partial t} dx \, \} dt.$$

Changing the order of integration, we obtain

$$\overline{S} = \int_{t_1}^{t_2} \int_{-\infty}^{\infty} \left( \frac{\hbar^2}{2m} \psi(x,t) \frac{\partial^2 \psi(x,t)}{\partial x^2} - U(x) \psi(x,t)^2 + E \psi(x,t)^2 \pm i\hbar \psi(x,t) \frac{\partial \psi(x,t)}{\partial t} \right) dx dt.$$

(91)

From the Ex. (91) follows the final form of the time-dependent average action for a chaotically wandering particle

$$\overline{S} = \int_{t_1}^{t_2} \int_{-\infty}^{\infty} \left( \frac{\hbar^2}{2m} \psi(x,t) \frac{\partial^2 \psi(x,t)}{\partial x^2} + \psi(x,t)^2 [E - U(x,t)] \pm i\hbar \psi(x,t) \frac{\partial \psi(x,t)}{\partial t} \right) dx dt. \quad (92)$$

The condition for the extremality of the average action (i.e., the functional) (92) requires that its first variation vanish [20]

$$\delta \overline{S} = \delta \int_{t_1}^{t_2} \int_{-\infty}^{\infty} \left( \frac{\hbar^2}{2m} \psi(x,t) \frac{\partial^2 \psi(x,t)}{\partial x^2} + \psi(x,t)^2 [E - U(x,t)] \pm i\hbar \psi(x,t) \frac{\partial \psi(x,t)}{\partial t} \right) dx dt = 0.$$

(93)

Let's find the extremal of functional (92), i.e., the function $\psi(x,t)$ for which the average action (92) takes an extreme value.

First, recall that an extremal of a functional of the form

$$S = \iint L\left( x, t, z, \frac{\partial z}{\partial x}, \frac{\partial z}{\partial t}, \frac{\partial^2 z}{\partial x^2}, \frac{\partial^2 z}{\partial t^2}, \frac{\partial^2 z}{\partial t \partial x} \right) dxdt, \text{ где } z = \psi(x,t), \tag{94}$$

defined by the Euler – Poisson equation [20, p.316]

$$L_z - \frac{\partial}{\partial x}\{L_p\} - \frac{\partial}{\partial t}\{L_g\} + \frac{\partial^2}{\partial x^2}\{L_r\} + \frac{\partial^2}{\partial t^2}\{L_t\} + \frac{\partial^2}{\partial x \partial t}\{L_s\} = 0, \tag{95}$$

where $L_z$ is the derivative of the Lagrangian $L$ with respect to $z = \psi(x,t)$; (96)

$L_p$ is the derivative of the $L$ with respect to $p = \dfrac{\partial z}{\partial x} = \dfrac{\partial \psi(x,t)}{\partial x}$;

$L_g$ is the derivative of the $L$ with respect to $g = \dfrac{\partial z}{\partial t} = \dfrac{\partial \psi(x,t)}{\partial t}$;

$L_r$ is the derivative of the $L$ with respect to $r = \dfrac{\partial^2 z}{\partial x^2} = \dfrac{\partial^2 \psi(x,t)}{\partial x^2}$;

$L_t$ is the derivative of the $L$ with respect to $t = \dfrac{\partial^2 z}{\partial t^2} = \dfrac{\partial^2 \psi(x,t)}{\partial t^2}$;

$L_s$ is the derivative of the $L$ with respect to $s = \dfrac{\partial^2 z}{\partial t \partial x} = \dfrac{\partial^2 \psi(x,t)}{\partial t \partial x}$,

wherein

$$\frac{\partial}{\partial x}\{L_p\} = L_{px} + L_{pz}\frac{\partial z}{\partial x} + L_{pp}\frac{\partial p}{\partial x} + L_{pg}\frac{\partial g}{\partial x} \tag{97}$$

is the total partial derivative with respect to $x$;

$$\frac{\partial}{\partial t}\{L_g\} = L_{gt} + L_{gz}\frac{\partial z}{\partial t} + L_{gp}\frac{\partial p}{\partial t} + L_{gg}\frac{\partial g}{\partial t}$$

is the total partial derivative with respect to $t$;

$$\frac{\partial^2}{\partial x^2}\{L_r\} = \frac{\partial^2 L_r}{\partial x^2} + L_{rz}\frac{\partial z}{\partial x^2} + L_{rp}\frac{\partial p}{\partial x^2} + L_{rg}\frac{\partial g}{\partial x^2}$$

is the full second partial derivative with respect to $x^2$;

$$\frac{\partial^2}{\partial t^2}\{L_t\} = \frac{\partial^2 L_t}{\partial t^2} + L_{tz}\frac{\partial z}{\partial t^2} + L_{tp}\frac{\partial p}{\partial t^2} + L_{tg}\frac{\partial g}{\partial t^2}$$

is the full second partial derivative with respect to $t^2$;

$$\frac{\partial^2}{\partial t \partial x}\{L_s\} = \frac{\partial^2 L_s}{\partial t \partial x} + L_{sz}\frac{\partial z}{\partial t \partial x} + L_{sp}\frac{\partial p}{\partial t \partial x} + L_{sg}\frac{\partial g}{\partial t \partial x}$$

is the total mixed partial derivative with respect to $t$ and $x$.

To determine the terms included in the Euler - Poisson equation (95), we use the integrand from the time-dependent average action (92)

$$L=\frac{\hbar^2}{2m}\psi(x,t)\frac{\partial^2\psi(x,t)}{\partial x^2}+\psi(x,t)^2[E-U(x)]\pm i\hbar\psi(x,t)\frac{\partial\psi(x,t)}{\partial t}. \qquad (98)$$

As a result of substituting the Lagrangian (98) into Ex.s (96) and (97), we obtain (99)

$$L_z=\frac{\hbar^2}{2m}\frac{\partial^2\psi(x,t)}{\partial x^2}+2\psi(x,t)[E-U(x)]+i\hbar\frac{\partial\psi(x,t)}{\partial t}; \qquad \frac{\partial^2}{\partial t^2}\{L_t\}=0;$$

$$\frac{\partial}{\partial x}\{L_p\}=0; \qquad \frac{\partial^2}{\partial x\partial t}\{L_s\}=0;$$

$$\frac{\partial}{\partial t}\{L_g\}=\pm 2i\hbar\frac{\partial\psi(x,t)}{\partial t}; \qquad \frac{\partial^2}{\partial x^2}\{L_r\}=2\frac{\hbar^2}{2m}\frac{\partial^2\psi(x,t)}{\partial x^2}.$$

Substituting Ex.s (99) into the Euler - Poisson equation (95), we obtain the desired equation for determining the extremal $\psi(x,t)$ of the average action (i.e., the functional) (92)

$$i\hbar\frac{\partial\psi(x,t)}{\partial t}=\frac{3\hbar^2}{2m}\frac{\partial^2\psi(x,t)}{\partial x^2}+2[E-U(x,t)]\psi(x,t). \qquad (100)$$

Generalization to three dimensions reduces to an increase in the number of integrations, and instead of Eq. (100), we obtain

$$\pm i\hbar\frac{\partial\psi(x,y,z,t)}{\partial t}=\frac{3\hbar^2}{2m}\left\{\frac{\partial^2\psi(x,y,z,t)}{\partial x^2}+\frac{\partial^2\psi(x,y,z,t)}{\partial y^2}+\frac{\partial^2\psi(x,y,z,t)}{\partial z^2}\right\}+2[E-U(x,y,z)]\psi(x,y,z,t) \qquad (101)$$

or in compact form

$$\pm i\hbar\frac{\partial\psi(\bar{r},t)}{\partial t}=\frac{3\hbar^2}{2m}\nabla^2\psi(\bar{r},t)+2[E-U(\bar{r},t)]\psi(\bar{r},t), \qquad (102)$$

where **r** is the radius vector with the beginning in the "center" of the investigated object ($r^2 = x^2 + y^2 + z^2$) (see Figure 1);

$\nabla^2=\frac{\partial^2}{\partial x^2}+\frac{\partial^2}{\partial y^2}+\frac{\partial^2}{\partial z^2}$ is the Laplace operator.

Ex. (102) will be called the **time-dependent stochastic Euler-Poisson equation.**

Taking into account Ex. (6), Eq. (102) can be represented as

$$\pm i\hbar \frac{\partial \psi(\bar{r},t)}{\partial t} = \frac{3\hbar^2}{2m}\nabla^2\psi(\bar{r},t) + 2 < T(\vec{r},t) > \psi(\vec{r},t), \qquad (103)$$

where $< T(\vec{r},t) >$ is the average kinetic energy of a chaotically wandering particle at a point with coordinates $x$, $y$, $z$. In this case, the given average kinetic energy of the particle can change at this point with time $t$.

*It is interesting to note that if equation (5a), originally written by Schrödinger in [1], is represented as*

$$\pm 2i\hbar \frac{\partial \psi}{\partial t} = \Delta\,\psi - 2V\,\psi,$$

*then it becomes clear that initially there was a sign* $\pm$ *in front of the left side of this equation, as in (103), and before the energy V there was a coefficient 2.*

The time-dependent stochastic Euler-Poisson's equation (102) allows finding the extremal $\psi(x,y,z,t)$ of the functional of the averaged action of a chaotically wandering particle (92). The square of the modulus of a given extremal $|\psi(x,y,z,t)|^2$ is a of the probability density function to find a particle at a point with coordinates $x$, $y$, $z$ at time $t$.

In other words, the stochastic Euler-Poisson equation (102) plays exactly the same role in the considered stochastic process (see Figures 1 and 2), which the time-dependent Schrödinger equation (5) performs in quantum mechanics.

### 3.4 Derivation of the generalized time-independent Schrödinger equation

The stochastic equation (102) derived in this work is intended to describe the dynamics of the averaged state of a three-dimensional random process in which a chaotically wandering particle participates. It is obvious that this equation is somewhat different from the usual form of the time-dependent Schrödinger equation (5)

$$i\hbar\frac{\partial\psi(\vec{r},t)}{\partial t}=-\frac{\hbar^2}{2m}\nabla^2\psi(\vec{r},t)+U(\vec{r},t)\,\psi(\vec{r},t). \qquad (104)$$

However, in the stationary case, when the wave function, its total mechanical energy and potential energy do not depend on time [i.e. when $\psi(x,y,z,t)=\psi(x,y,z)$, $E(t)=E$ и $U(x,y,z,t)=U(x,y,z)$ ], stochastic equation (102) takes the form

$$-\frac{3}{2}\frac{\hbar^2}{2m}\nabla^2\psi(\vec{r})+U(\vec{r})\,\psi(\vec{r})=E\,\psi(\vec{r}). \qquad (105)$$

This equation is reduced to the form of the time-independent Schrödinger equation.

*Eq. (104) can be obtained from the condition of extremality of the time-independent average action (i.e., functional)*

$$\overline{S}=\int_{-\infty}^{\infty}\left(\frac{\hbar^2}{2m}\psi(x)\frac{\partial^2\psi(x)}{\partial x^2}+\psi(x)^2[E-U(x)]\right)dx, \qquad (105a)$$

*which is obtained from time-dependent functional (92) under the condition of stationarity. Indeed, having performed actions similar to (93) – (100) using the integrand from (104a), we obtain for the three-dimensional case equation (104) for the extremal* $\psi(\vec{r})$.

To show how the form of the time-independent Schrödinger equation is obtained, we introduce the notation:

$$\varepsilon = E/m \qquad (106)$$

- let’s call this massless quantity the "total mechanical energiality" of the particle;

$$u(x,y,z) = U(x,y,z)/m \qquad (107)$$

- let’s call this massless quantity the "potential energiality" of the particle.

Taking into account Ex.s (106) and (107), Eq. (105) can be represented in the form

$$\frac{3\hbar^2}{4m}\nabla^2\psi(x,y,z)+\ m[\varepsilon-u(x,y,z)]\,\psi(x,y,z)=0. \tag{108}$$

We divide both sides of the Eq. (108) by $\hbar$

$$\frac{3\hbar}{4m}\nabla^2\psi(x,y,z)+\ \frac{m}{\hbar}[\varepsilon-u(x,y,z)]\,\psi(x,y,z)=0, \tag{109}$$

and take into account that there is a connection between the ratio $\hbar/m$ and the main characteristics of a stationary random process (61)

$$\eta_p=\frac{2\sigma_{pr}^2}{\tau_{pr}}=\frac{\hbar}{m}, \tag{110}$$

where for the three-dimensional case

$$\sigma_{pr}=\frac{1}{3}\sqrt{\sigma_{px}^2+\sigma_{py}^2+\sigma_{pz}^2} \tag{111}$$

is the standard deviation of a chaotically wandering particle from the conditional center of the stochastic system (Figures 1 and 2);

$$\tau_{pr}=\frac{1}{3}\left(\tau_{px}+\tau_{py}+\tau_{pz}\right) \tag{112}$$

is the averaged autocorrelation interval of the considered three-dimensional stationary random process (Figure 2).

Taking into account (110), Ex. (109) takes the form

$$\frac{3}{4}\eta_p\nabla^2\psi(x,y,z)+\ \frac{1}{\eta_p}[\varepsilon-u(x,y,z)]\,\psi(x,y,z)=0.$$

Let's divide both sides of this expression by $\frac{3}{4}\eta_p$ and, as a result of simple transformations, we obtain

$$\nabla^2\psi(x,y,z)+\frac{2}{\eta_{p1}^2}[\varepsilon-u(x,y,z)]\,\psi(x,y,z)=0, \tag{113}$$

where

$$\eta_{p1}=\sqrt{\frac{3}{2}\eta_p}=\sqrt{\frac{3}{2}\frac{2\sigma_{pr}^2}{\tau_{pr}}}=\sqrt{6}\,\frac{\sigma_{pr}^2}{\tau_{pr}} \tag{114}$$

is the scale parameter of the investigated three-dimensional random process (i.e., chaotic wanders of a particle in the vicinity of the "center", see Figures 1, 2 & 3).

The form of the obtained Ex. (113) completely coincides with the form of the time-independent Schrödinger equation

$$\nabla^2\psi(x,y,z)+\frac{2m}{\hbar^2}[E-U(x,y,z)]\,\psi(x,y,z)=0, \qquad (115)$$

which, taking into account (106) and (107), can be represented as

$$\nabla^2\psi(x,y,z)+\frac{2m^2}{\hbar^2}[\varepsilon-u(x,y,z)]\,\psi(x,y,z)=0. \qquad (116)$$

Eq.s (113) and (116) have the same form

$$\nabla^2\psi(x,y,z)+\text{const}\cdot[\varepsilon-u(x,y,z)]\,\psi(x,y,z)=0.$$

Therefore, Ex. (113) we will be called the **massless generalized time-independent Schrödinger equation**.

Meanwhile, stochastic equation (113) has two tangible advantages over the Schrödinger equation (116):

1] Eq. (116) is a special case of Eq. (113). When studying the averaged behavior of a particle of subatomic scale (for example, an electron), the average characteristics of its chaotic behavior $\sigma_{pr} = \sigma_{er}$ and $\tau_{pr} = \tau_{er}$ included in Ex. (114) can be such that the relation is performed

$$\frac{2}{\eta_{e1}^2}\approx\frac{2m_e^2}{\hbar^2}. \qquad (117)$$

In this case, Eq.s (113) and (116) almost completely coincide.

When considering the chaotic behavior of a microparticle (for example, the nucleus of a biological cell), the averaged parameters $\sigma_{pr} = \sigma_{nr}$ и $\tau_{pr} = \tau_{nr}$ can also be estimated (measured) and substituted into Ex. (114). In this case, the generalized time-independent Schrödinger equation (113) will describe a discrete (quantized) set of possible averaged states of the nucleus of a biological cell.

Equation (113) is also suitable for describing the average states of chaotic oscillations or displacements of the center of mass: the yolk in an egg, a criminal in a prison cell, a nucleus in the bowels of the planet, etc.

2] The generalized time-independent Schrödinger equation (113) follows mathematically correctly and logically from the stochastic Euler-Poisson's equation (102). Whereas it is quite obvious that it is impossible to directly obtain the time-independent Schrödinger equation (115) from the time-dependent Schrödinger equation (104). This is achieved, however, in a very strange way for ordinary consciousness by representing the wave function $\psi(x,y,z,t)$ in the form of a complex function

$$\psi(x,y,z,t) = \psi(x,y,z)\ exp\{iEt/\hbar\}, \qquad (118)$$

which is justified by the formal concepts of de Broglie's matter waves, which are not experimentally observed.

Eq. (113) has an advantage over Eq. (115) in that it is applicable not only for describing the quantum effects of the microworld, but also for the averaged behavior of macroscopic stochastic objects under similar conditions, and for which, as a result of statistical processing, their chaotic behavior can to obtain the characteristics $\sigma_{pr}$ and $\tau_{pr}$ . Such stochastic objects (regardless of their scale) can include, for example, an electron in a hydrogen atom, a wandering nucleus of a biological cell, a quivering yolk in an egg, a fly in a glass jar, the tip of a tree branch under the influence of gusts of wind, the center of mass of a mobile embryo in the womb, an oscillating core in the interior of a star or planet, etc.)

In other words, Eq. (113) is suitable for describing the averaged state of any stochastic object that constantly chaotically wanders (trembles, vibrates, moves) in the vicinity of the conditional center, in such a way that its total mechanical energy (more precisely, energiality) remains unchanged.

This is the novelty of this work, and not only that it was possible to correctly derive the stochastic equation (113) similar to the time-independent Schrödinger equation (115). This is the novelty of this work, and not only that it was possible to correctly derive the stationary stochastic equation (113) exactly similar to the time-independent Schrödinger equation (115).

### 3.5 Stochastic quantum operators

Let us show how operators are obtained in the stochastic quantum mechanics proposed in this article. To do this, let's return to the model shown in Figures 1 and 2. During the chaotic movement of the particle in the vicinity of the conditional center, it constantly changes the direction of its movement (Figures 1 and 2). Therefore, the particle at each moment of time has an angular momentum

$$\vec{L} = \vec{r} \times \vec{p}, \tag{119}$$

where $\vec{r}$ is the radius vector from the conditional center to the particle (see Fig. 2);

$\vec{p} = m_n \vec{v}$ – instantaneous value of particle’s momentum.

Let’s represent the vector Eq. (63) in the component form

$$L_x = yp_z - zp_y, \quad L_y = zp_x - xp_z, \quad L_z = xp_y - yp_x. \tag{120}$$

Let's average these components

$$\overline{L}_x = y\overline{p_z} - z\overline{p_y}, \quad \overline{L}_y = z\overline{p_x} - x\overline{p_z}, \quad \overline{L}_z = x\overline{p_y} - y\overline{p_x}. \tag{121}$$

We use the coordinate representation of the average $x$-component of the momentum (81) (122)

$$\overline{p_x} = \int_{-\infty}^{\infty} \rho(p_x) p_x dp_x = -i\hbar \int_{-\infty}^{\infty} \psi(x) \frac{\partial\, \psi(x)}{\partial x} dx = \int_{-\infty}^{\infty} \psi(x) \left( -i\hbar \frac{\partial}{\partial x} \right) \psi(x) dx \equiv \left( -i\hbar \frac{\partial}{\partial x} \right) \int_{-\infty}^{\infty} \psi(x)\psi(x) dx.$$

Similarly

$$\overline{p_y} = \int_{-\infty}^{\infty} \rho(p_y) p_y dp_y \equiv \left( -i\hbar \frac{\partial}{\partial y} \right) \int_{-\infty}^{\infty} \psi(y)\psi(y) dy, \tag{123}$$

$$\overline{p_z} = \int_{-\infty}^{\infty} \rho(p_z) p_z dp_z \equiv \left( -i\hbar \frac{\partial}{\partial z} \right) \int_{-\infty}^{\infty} \psi(z)\psi(z) dz. \tag{124}$$

Let us take into account that, for example, in Ex. (122)

$$\int_{-\infty}^{\infty} \psi(x)\psi(x) dx = \int_{-\infty}^{\infty} \rho(x) dx = 1, \tag{125}$$

Then identities (122), (123) and (124) turn out to be equivalent to the operators of the components of the momentum vector

$$\hat{p}_x = \frac{\hbar}{i}\frac{\partial}{\partial x}, \quad \hat{p}_y = \frac{\hbar}{i}\frac{\partial}{\partial y}, \quad \hat{p}_z = \frac{\hbar}{i}\frac{\partial}{\partial z}. \tag{126}$$

Let's substitute Ex.s (122) through (124) into Ex. (121) and take into account (125) and (126). As a result, we obtain the operators of the angular momentum components [73]

$$\hat{L}_x = \frac{\hbar}{i}\left(y\frac{\partial}{\partial z} - z\frac{\partial}{\partial y}\right), \quad \hat{L}_y = \frac{\hbar}{i}\left(z\frac{\partial}{\partial x} - x\frac{\partial}{\partial z}\right), \quad \hat{L}_z = \frac{\hbar}{i}\left(x\frac{\partial}{\partial y} - y\frac{\partial}{\partial x}\right).$$

Let's divide both sides of these expressions by the $m|\vec{r}\,|^2$

$$\begin{aligned}
\frac{\hat{L}_x}{m|\vec{r}\,|^2} &= \frac{\hbar}{m|\vec{r}\,|^2\, i}\left(y\frac{\partial}{\partial z} - z\frac{\partial}{\partial y}\right),\\
\frac{\hat{L}_y}{m|\vec{r}\,|^2} &= \frac{\hbar}{m|\vec{r}\,|^2\, i}\left(z\frac{\partial}{\partial x} - x\frac{\partial}{\partial z}\right),\\
\frac{\hat{L}_z}{m|\vec{r}\,|^2} &= \frac{\hbar}{m|\vec{r}\,|^2\, i}\left(x\frac{\partial}{\partial y} - y\frac{\partial}{\partial x}\right).
\end{aligned} \tag{127}$$

Taking into account (61), we have massless stochastic quantum operators

$$\omega_x = \frac{\eta_p}{i|\vec{r}\,|^2}\left(y\frac{\partial}{\partial z} - z\frac{\partial}{\partial y}\right), \quad \hat{\omega}_y = \frac{\eta_p}{i|\vec{r}\,|^2}\left(z\frac{\partial}{\partial x} - x\frac{\partial}{\partial z}\right), \quad \hat{\omega}_z = \frac{\eta_p}{i|\vec{r}\,|^2}\left(x\frac{\partial}{\partial y} - y\frac{\partial}{\partial x}\right), \tag{128}$$

where $\hat{\omega}_x$, $\hat{\omega}_y$, $\hat{\omega}_z$ are the components of the stochastic operator of the angular velocity of a chaotically wandering particle, since

$$\vec{\omega} = \frac{\vec{L}}{m|\vec{r}\,|^2} = \frac{\vec{r}\times\vec{v}}{|\vec{r}\,|^2}. \tag{129}$$

In a spherical coordinate system, massless stochastic operators (128) have the form

$$\hat{\omega}_x = \frac{\eta_p}{i\,|\vec{r}\,|^2}\left(\sin\varphi\frac{\partial}{\partial\theta} - ctg\theta\,\cos\varphi\frac{\partial}{\partial\varphi}\right),$$

$$\hat{\omega}_y = \frac{\eta_p}{i\,|\vec{r}\,|^2}\left(\cos\varphi\frac{\partial}{\partial\theta} - ctg\theta\sin\varphi\frac{\partial}{\partial\varphi}\right), \qquad (130)$$

$$\hat{\omega}_z = \frac{\eta_p}{i\,|\vec{r}\,|^2}\frac{\partial}{\partial\varphi}.$$

The stochastic quantum operator of the squared of the angular velocity of a chaotically wandering particle is

$$\hat{\omega}^2 = \hat{\omega}_x^2 + \hat{\omega}_y^2 + \hat{\omega}_z^2 = -\frac{\eta_p^2}{|\vec{r}\,|^4}\nabla^2_{\theta,\varphi}, \qquad (131)$$

where [73]

$$\nabla^2_{\theta,\varphi} = \frac{1}{\sin\theta}\frac{\partial}{\partial\theta}\left(\sin\theta\frac{\partial}{\partial\theta}\right) + \frac{1}{\sin^2\theta}\frac{\partial^2}{\partial\varphi^2}. \qquad (132)$$

Now let's divide both sides of the Ex.s (126) by the particle mass $m$

$$\frac{\hat{p}_x}{m} = \frac{\hbar}{im}\frac{\partial}{\partial x}, \quad \frac{\hat{p}_y}{m} = \frac{\hbar}{im}\frac{\partial}{\partial y}, \quad \frac{\hat{p}_z}{m} = \frac{\hbar}{im}\frac{\partial}{\partial z}. \qquad (133)$$

As a result, we obtain stochastic quantum operators of the components of the velocity vector

$$\hat{v}_x = \frac{\eta_p}{i}\frac{\partial}{\partial x}, \quad \hat{v}_y = \frac{\eta_p}{i}\frac{\partial}{\partial y}, \quad \hat{v}_z = \frac{\eta_p}{i}\frac{\partial}{\partial z}. \qquad (134)$$

Many other stochastic quantum operators, analogous to quantum mechanical operators, can be obtained in a similar way.

### 3.6 Quantized states of the nucleus of a biological cell

Stochastic quantum physics, proposed in this article, is suitable for describing quantum effects not only at the level of elementary particles, but also at the micro- and macro-levels of the world around us. As an example, we use Eq. (113) to study the average behavior of a chaotically fluctuating nucleus of a biological cell (hereinafter «cell nucleus» or «*c*-nucleus») (see Figure 5).

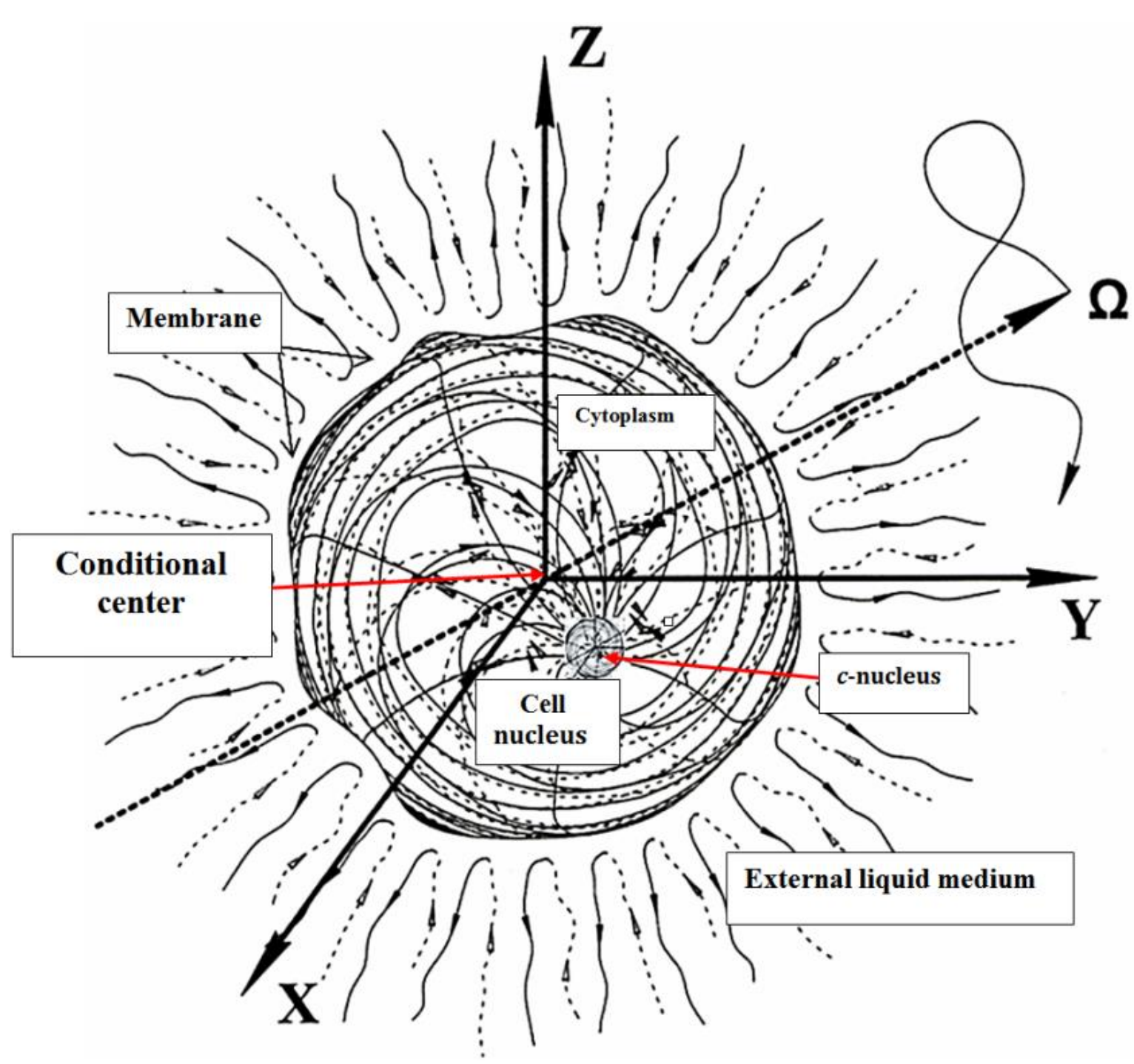

**Fig.5:** A simplified model of a eukaryotic biological cell with a pronounced cell nucleus, which continuously wanders (fluctuates) in the vicinity of the conditional center, so that its total mechanical energy $E$ always remains constant (E = *const*)

Consider a living eukaryotic biological cell in the period between its division (i.e., in a state of interphase).

Let the state of the interphase of the biological cell under consideration continue for the entire period of observation of the chaotically oscillating cell nucleus (hereinafter the *c*-nucleus). In this case, the total mechanical energiality of the *c*-nucleus remains constant

$$\varepsilon_n = t_n(x,y,z,t) + u_n(x,y,z,t) = const, \tag{136}$$

where $\varepsilon_n = E_n(x,y,z,t)/m_n$ is the total mechanical energiality of the *c*-nucleus (here $m_n$ is the mass of the k-nucleus);

$t_n(x,y,z,t) = T_n(x,y,z,t)/m_n$ is the kinetic energiality of the *c*-nucleus;

$u_n(x,y,z,t) = U_n(x,y,z,t)/m_n$ is the potential energiality of the *c*-nucleus.

The considered random process with the participation of the $c$-nucleus completely corresponds to the stochastic model described in § 2.1, and condition (136) coincides with condition (6). Therefore, to describe the average states of a chaotically oscillating $c$-nucleus, the massless generalized time-independent Schrödinger equation (113) can be used

$$\nabla^2\psi(x,y,z)+\frac{2}{\eta_{n1}^2}[\varepsilon_n-u_n(x,y,z)]\,\psi(x,y,z)=0, \qquad (137)$$

where $\eta_{n1}=\sqrt{\frac{3}{2}\frac{2\sigma_{nr}^2}{\tau_{nr}}}$ is scale parameter; (138)

$$\sigma_{nr}=\frac{1}{3}\sqrt{\sigma_{nx}^2+\sigma_{ny}^2+\sigma_{nz}^2} \qquad (139)$$

– is the standard deviation of a chaotically oscillating $c$-nucleus from a conventional center (see Figure 5);

$$\tau_{nr}=\frac{1}{3}\left(\tau_{nx}+\tau_{ny}+\tau_{nz}\right) \qquad (140)$$

– is the averaged autocorrelation interval of the considered three-dimensional stationary random process, in which a chaotically oscillating $c$-nucleus participates.

Let us consider the case when the elastic tensions of the cytoplasm $\sigma_v$, surrounding the $c$-nucleus, increase on average in proportion to the distance of the $c$-nucleus from the conventional center

$$\sigma_v(x)\approx k_u r, \qquad (141)$$

where $k_u = K_u/m_n$ is the massless coefficient of elastic tension of the cytoplasm;

$K_u$ is the force constant of the cytoplasm;

$r=\sqrt{x^2+y^2+z^2}$ is distance from the conventional center to the $c$-nucleus.

In this case, the averaged potential energiality of the $c$-nucleus can be approximately represented in the form

$$u_n(r)\approx\int k_{ux}\,r\,dr=\frac{1}{2}k_u r^2. \qquad (142)$$

Substituting Ex. (142) into Eq. (137), we obtain the equation of an isotropic three-dimensional harmonic oscillator known in quantum mechanics [74]

$$\nabla^2\psi(r) + \frac{2}{\eta_{nr1}^2}[\varepsilon_n - \frac{k_u r^2}{2}]\psi(r) = 0, \quad (143)$$

where
$$\nabla^2 = \frac{1}{r^2}\frac{\partial}{\partial r}\left(r^2\frac{\partial}{\partial r}\right) + \frac{\nabla^2_{\theta,\varphi}}{r^2} \quad (144)$$

is the Laplace operator in spherical coordinates, while the operator $\nabla^2_{\theta,\varphi}$ is given by Ex. (132);

The solutions to Eq. (143) are wave functions [74]

$$\psi_{klm}(r,\theta,\varphi) = R_k(r)Y_{lm}(\theta,\varphi) = \sqrt{\sqrt{\frac{2}{\pi}\left(\frac{\sqrt{k_u}}{2\eta_{n1}}\right)^3}\frac{2^{k+2l+3}k!}{(2k+2l+1)!!}\left(\frac{\sqrt{k_u}}{2\eta_{n1}}\right)^l}\, r^l \exp\left\{-\frac{\sqrt{k_u}\,r^2}{2\eta_{n1}}\right\}\times$$
$$\times L_l^{(l+1/2)}\left(2\sqrt{\frac{\sqrt{k_u}}{2\eta_{nx1}}}r^2\right)Y_{lm}(\theta,\varphi), \quad (145)$$

where $L_l^{(l+1/2)}\left(2\sqrt{\frac{\sqrt{k_u}}{2\eta_{nx1}}}r^2\right)$ are a generalized Laguerre polynomials;

$Y_{lm}(\theta,\varphi) = (-1)^m\left[\frac{(2l+1)(l-m)!}{4\pi(l+m)!}\right]^{\frac{1}{2}} e^{im\varphi}P_{lm}(\cos\theta)$ are a spherical harmonic functions;

$P_{lm}(\cos\theta) = \frac{d}{2^l l!}\left(1-\xi^2\right)^{m/2}\frac{d^{l+m}}{d\xi^{l+m}} + \left(\xi^2-1\right)^l$ are associated Legendre functions;

$\xi = \cos\theta$;

$l$ is the azimuthal quantum number;

$m$ is the peripheral quantum number.

*In atomic quantum physics, the number m is called the "magnetic quantum number", but this name is not suitable for stochastic quantum physics. Therefore, in this article, the number m is proposed to be called the “peripheral quantum number”.*

*As you know, there is also a fourth spin quantum number s, which in the case under consideration, seemingly, is associated with one of the two possible*

*directions of rotation of the cytoplasm inside a biological cell. However, this process is not covered in this article.*

The wave functions (145) correspond to the eigenvalues of the total mechanical energiality of the $c$-nucleus [74]

$$\varepsilon_{nkl} = \eta_{nr1}\sqrt{k_{ur}}\left(2k + l + \frac{3}{2}\right) = \eta_{nr1}\sqrt{k_{ur}}\left(N + \frac{3}{2}\right), \quad \text{where } N = 2k + l. \qquad (146)$$

The squares of the modulus of wave functions (145) $|\psi_{klm}(r,\theta,\varphi)|^2$ (i.e., the PDF of the possible location of the $c$-nucleus inside the biological cell) at $\varphi = 0$ and different values of the quantum numbers $k$, $l$ and $m$ are shown in Figure 6.

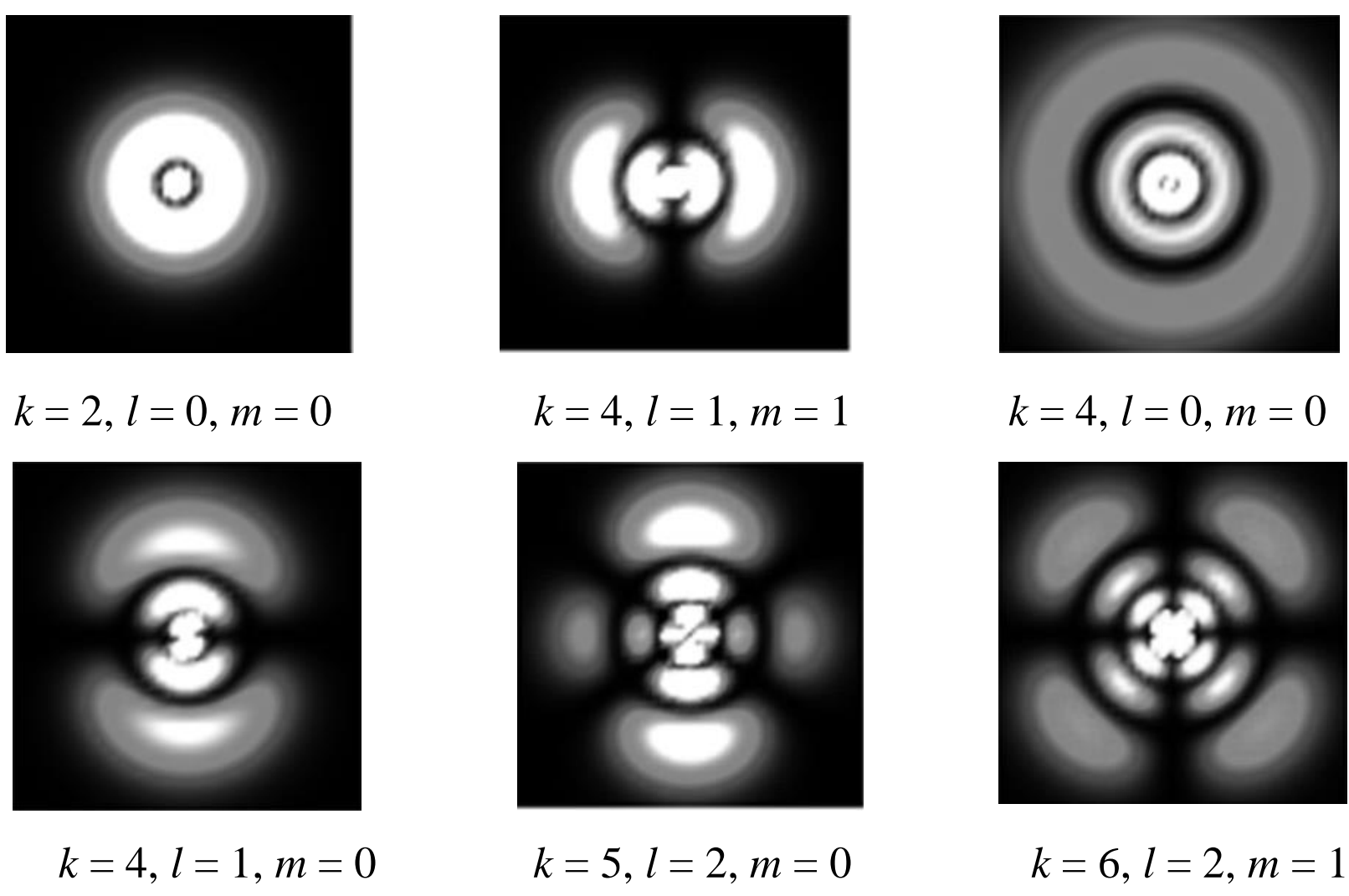

**Fig. 6:** The probability density function (PDF) $|\psi_{klm}(r,\theta,\varphi)|^2$ possible locations of the $c$-nucleus at $\varphi = 0$ and different values of the quantum numbers $k$, $l$ and $m$. The lighter the spot, the more likely a $c$-nucleus will appear in this area. Calculations were performed using Ex. (145) [74] and presented on the web page: Spherical_Harmonic_Orbitals.png

The Figure 6 shows that each set of three quantum numbers $k$, $l$, and $m$ corresponds to a unique spatial configuration of the average state of the chaotic wandering (oscillating or trembling) of the $c$-nucleus. This state is determined by the PDF of the place of possible appearance of the $c$-nucleus (more precisely, its center of mass), inside the biological cell.

*As is known, there is also a fourth spin quantum number s, which in the case under consideration, seemingly, is associated with one of two possible directions of rotation of the cytoplasm inside the biological cell. However, this process is not covered in this article.*

To experimentally fix one of the spatial configurations of the averaged behavior of the *c*-nucleus, it is necessary:

- to ensure the absence of a tangible influence of external and internal force factors on the biological cell for the entire period of observation of the *c*-nucleus.

- make a video recording of the chaotic behavior of the *c*-nucleus for a long period of time;

- take into account that the biological cell as a whole can participate in complex Brownian and/or thermal motion; these movements must be eliminated physically or excluded by software;

- take into account that the *c*-nucleus can change its shape, and its internal content (karyoplasm, chromatin etc.) can change over time; this leads to blurring of the boundaries of the spatial configuration of the averaged state of trembling of the given organelle. Therefore, it is necessary to monitor not the behavior of the entire cell nucleus, but the chaotic movement of only its center of mass. In other words, it is needful to programmatically identify the center of mass of the cell nucleus (i.e., the *c*-nucleus) and monitor only it's chaotic motion.

- digitized, software-cleaned and mathematically processed video recording of the chaotic behavior of the *c*-nucleus (i.e., the its center of mass) need to reproduce at high speed, with the display of this highly accelerated process on the computer monitor. The speed of reproduction of the motion of the *c*-nucleus should be so fast that this point is "smeared" over the entire observation area.

- if all the above actions can be performed with a sufficiently high resolution of the video equipment, as cleanly as possible and with the exclusion of various interfering factors, then in accordance with the hypothesis set out in this article, the configuration of dark and light spots should be revealed on the monitor screen (similar to one of the spot configurations shown in Figure 7). In this case, a dark spot on the monitor screen should mean that the $c$-nucleus appeared in this place more often than in the place where the light spot was formed.

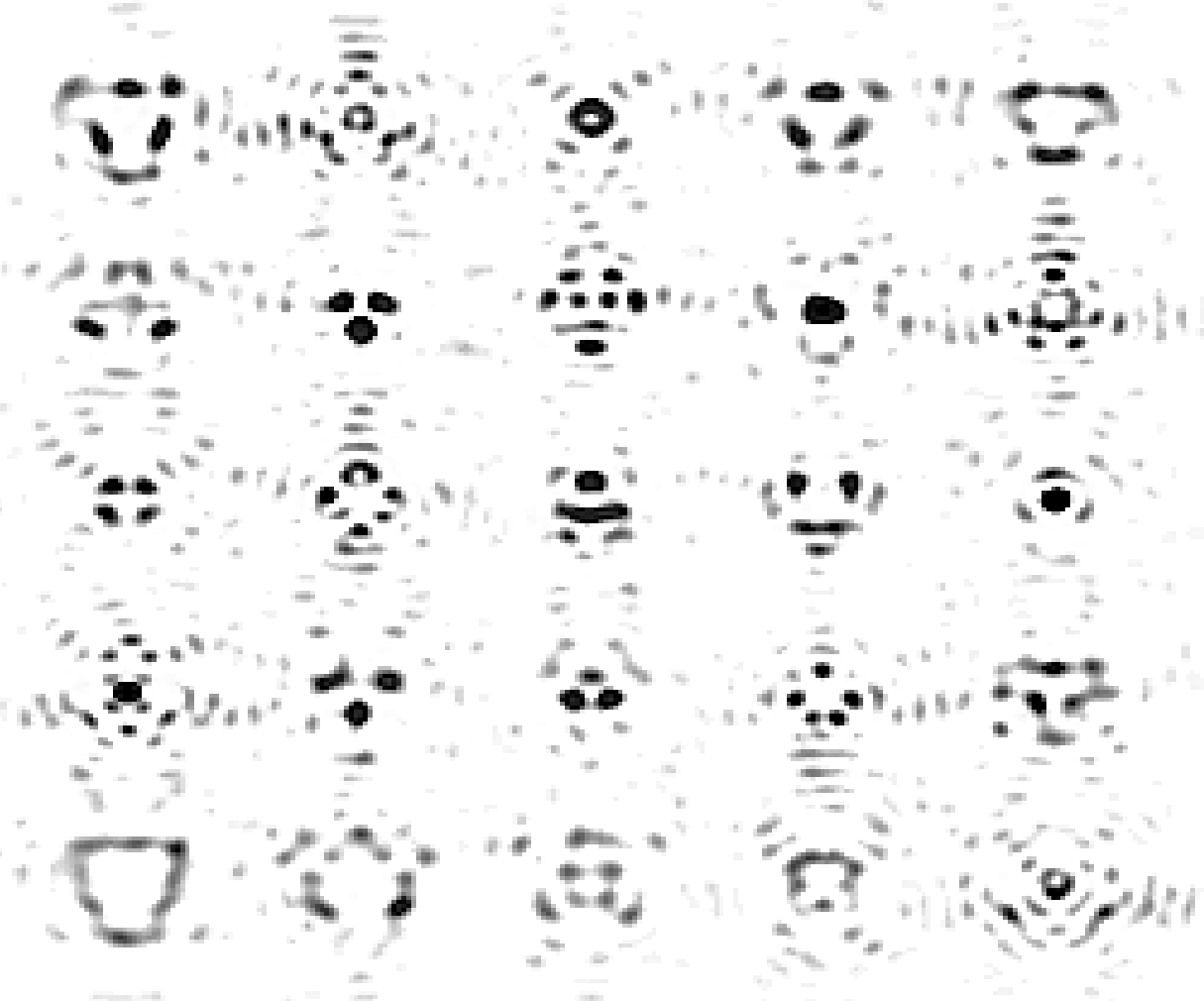

**Fig. 7:** Examples of possible configurations of dark-light spots that can be revealed as a result of averaging the chaotic movement (tremor) of the $c$-nucleus inside a biological cell. These spot configurations can correspond to eigenwave functions $\psi_{klm}(r,\theta,\varphi)$ (145) with different sets of three quantum numbers $k$, $l$, and $m$

It is possible that the configuration of these dark-light spots will to correspond to one of the eigenwave functions $\psi_{klm}(r,\theta,\varphi)$ (more precisely, PDF $|\psi_{klm}(r,\theta,\varphi)|^2$) of the isotropic three-dimensional quantum harmonic oscillator (see light-dark spots shown in Figure 6).

Using sound vibrations with a resonant frequency (i.e., with a frequency close to the natural frequency of vibrations of the considered isotropic three-dimensional quantum harmonic oscillator)

$$f_{0x} \approx \sqrt{\frac{3}{2}} \frac{1}{2\pi\tau_{nr}}, \tag{147}$$

we can to influence the cell nucleus, thereby changing the average state of the chaotic wandering of its center of mass (i.e. $c$-nucleus). After that, all the above steps should be repeated to identify a different configuration of dark-light spots, corresponding to a more excited state of the $c$-nucleus.

It is possible that these oscillations must be excited simultaneously from two opposite ends of each of the three mutually perpendicular directions. This will make it possible to exclude the possibility of translational displacement of the $c$-nucleus due to unilateral action.

If the principles of the stochastic quantum physics of biological cell organelles, described above, are correct, then the quantum transition of the $\underline{c}$-nucleus from one average state to its other state (i.e., a jump-like change in the configuration of dark-light spots on the monitor screen) should occur when transferring to a given $c$-nucleus additional portion of the total mechanical energiality

$$\Delta\varepsilon_n = \varepsilon_{nk} - \varepsilon_{nk+j}\,, \tag{148}$$

where $\varepsilon_{nk}$ is the total mechanical energiality of the $c$-nucleus in the state $k$;

$\varepsilon_{nk+\mathrm{j}}$ is the total mechanical energiality of the $c$-nucleus in the state $k + j$ (here $j$ = 1,2,3, ...).

In this respect, the stochastic quantum biophysics ($\sim 10^{-3}$cm) should not differ from the quantum physics of elementary particles ($\sim 10^{-13}$cm). The difference lies only in the scales of the processes under consideration, which differ from each other by about 10 orders of magnitude. Meanwhile, there is no doubt in the modern scientific community that quantum mechanics is applicable to describe processes on intermediate scales, i.e. at the molecular level ($\sim 10^{-8}$cm) [75,76]. For example, quantum-mechanical methods describe oscillations of the atomic lattice (phonons), molecular vibrations, etc.

## 4 DISCUSSION

The article considers a stationary random (stochastic) process associated with the chaotic motion of a particle in the vicinity of the conditionally selected center of a given stochastic system (Figure 1 or 5). Based on a detailed consideration of this process, we obtained stochastic Euler-Poisson's equation (102) for the extremal of the averaged functional of the action of a randomly wandering particle (92).

In the stationary case, when the wave function, its total mechanical energy and potential energy do not depend on time [i.e. when $\psi(x,y,z,t) = \psi(x,y,z)$, $E(t) = E$ and $U(x, y, z, t) = U(x, y, z)$ ], the stochastic Euler-Poisson equation (102) is reduced to the stationary stochastic Eq. (113) exactly similar to the time-independent Schrödinger equation (115).

Eq. (113) equally well describes the discrete sets of the averaged behavior of an electron behavior in a potential well, a nucleus in the cytoplasm of a biological cell, the center of the embryonic mass in the womb, a nucleus in the bowels of the planet, a fly in a bank, etc. All these stable stochastic processes have the possibility of transition from one stationary standing in another stationary standing with the absorption or release of a specific portion of the total mechanical energiality.

In this way, together with the derivation of the massless generalized time-independent Schrödinger equation (113), we come to the realization that quantum transitions are inherent not only to objects of atomic and subatomic scale, but also manifest themselves at all levels of the organization of being.

This is easy to verify, for example, by referring to a fly constantly flying in a large glass jar. With the help of a video camera, you can record its chaotic movements for a long time. If you then scroll through the video at a very high speed, you will not see a fly on the screen, but a stable blurry spot that reflects the probability density of the location of the fly. It should be expected that if the fly is not disturbed by anything, then the blurry spot will resemble a Gaussian probability distribution density with the greatest black in the center of the glass jar. However, if the fly is somehow influenced by an energy transfer, for example, by heat or by ultrasound with a certain frequency, then the average behavior of the fly can be changed dramatically (in discrete steps). In this case, a blurry spot can change the configuration to an averaged ring or an averaged eight, etc. (Of course, no fly should be made to suffer from such experiments.) The embryo in the womb, the core in the bowels of the planet, and many other similar objects will behave in the same way for long periods of time. It is precisely these different probabilistic configurations with different energy levels that are described by the massless generalized time-independent Schrödinger equation (113) derived in this article.

The center of the embryonic mass in the womb, and the core in the bowels of the planet, and many other similar objects whose behavior is studied over fairly long periods of time, will behave in the same way.

The approach proposed in this paper makes it possible to derive the basic equations of nonrelativistic stochastic quantum physics (102) and (113) based on principles that are fundamentally different from the ideological foundations of the Copenhagen and Many-Worlds interpretations of quantum mechanics. (For example, in this article, the wandering particle under study has a chaotic trajectory and specific dimensions.) However, the mathematical apparatus of quantum mechanics, created by great scientists, remained virtually unchanged, but its logical foundations becomes thereby much clearer.

In a similar way, all the basic equations of quantum field theory can be obtained: the Clen-Gordon equation, Dirac equations, Maxwell equations, etc. Their derivation algorithm is similar to the approach given in this article:

1) express the deterministic action of the particle;

2) find the mean of this action,

3) all the averaged terms in the integrand of the averaged action are represented through the probability density functions $\rho(x)$ and/or $\rho(p_x)$;

4) switch all terms of the Lagrangian of the averaged action to a coordinate representation or a momentum representation;

5) determine the equation for the extremal of the resulting functional (averaged action) through methods of the calculus of variations.

## 5 CONCLUSIONS

The meaning of the derivation of stochastic Eq.s (102) and (113) presented here is as follows:

- It becomes clear to which phenomena of the micro- and macrocosm these equations relate, what are the boundaries and conditions of their application.
- There is no need to apply either the Heisenberg “uncertainty principle” or the de Broglie concept of “matter waves”, since the procedure (45) through (48) is used to derive the stochastic Eq.s (102) and (113), and this procedure is completely analogous to the transition from the coordinate submission to impulse, and vice versa. However, this procedure is obtained only on the basis of an analysis of the properties of a stationary random process, without involving the above hypotheses. When solving this problem, an intermediate result was obtained: a procedure for obtaining the probability density function of the derivative of the nth order for an $n$-fold differentiable stationary random process was determined. This result can be used in various branches of statistical physics.
- The ratio $\hbar/m$ (“reduced Planck's constant” divided by mass) is determined through the variance and correlation time of the investigated stationary stochastic process (114). Consequently, the generalized stochastic equations (102) and (113) do not contain the particle "mass" $m$, and therefore it is necessary to introduce an additional dimensional constant - the reduced Planck's constant. “Mass” is (according to the author) one of the “darkest” dimensional quantities in modern physics (see Chapter 7 in Gauchman, 2008 [21]). There is no doubt that the concept of "mass" should be absent in the final theory, and this article is one of the steps to eradicate this concept from scientific concepts of nature.

- The volume and trajectory of the wandering particle are returned to consideration. Together with them, the physics of the microworld again finds its usual logical "ground underfoot."
- We hope that if this work is carefully analyzed and accepted by the scientific community, this will not only allow us to calculate the probabilistic results of complex chaotic processes in both the microcosm and the macrocosm, but also to analyze the inner essence of these processes, as suggested by Albert Einstein. That is, now we can “Think and calculate”.

## 6 ACKNOWLEDGEMENTS

I wish to thank D. Reid and Dr. V. A. Lukyanov for valuable comments made during the preparation of this article. I also express gratitude to my mentors, Dr. A.A. Kuznetsov and Dr. A.I. Kozlov. Great help in the discussion of this article was provided by Dr. A.A. Rukhadze, Dr. A.M. Ignatov and Academician of the Russian Academy of Natural Sciences G.I. Shipov. Thanks a lot for help and support Dr. T.S. Levy, A. N. Maslov and A. Yu. Bolotov. The review and comments by Dr. L. S. F. Olavo (*University of Brasília*) and Dr. D. Edwards (*University of Georgia, USA)* was helpful to the author.

## DATA AVAILABILITY

Data confirming the results of this study can be obtained from the author of this article upon request.